\documentclass[preprint]{elsarticle}

\usepackage{lineno,hyperref}
\modulolinenumbers[5]
\biboptions{sort&compress}

\usepackage{colortbl}
\usepackage{etoolbox}
\usepackage[table]{xcolor}

\def\prd{{\it Phys. Rev.} D~}

\def\apjl{{\it Astrophys. J.} Lett~}
\def\apj{{\it Astrophys. J.}}

\def\PRD{{\it Phys. Rev.} D~}
\def\prl{{\it Phys. Rev.} Lett~}

\def\mnras{{\it MNRAS~}}

\def\apj{{\it Astrophys. J.}}
\def\nat{{\it Nature}}

\def\lesssim{\mathrel{\hbox{\rlap{\hbox{\lower4pt\hbox{$\sim$}}}\hbox{$<$}}}}
\def\gtrsim{\mathrel{\hbox{\rlap{\hbox{\lower4pt\hbox{$\sim$}}}\hbox{$>$}}}}
\def\alt{\mathrel{\hbox{\rlap{\hbox{\lower4pt\hbox{$\sim$}}}\hbox{$<$}}}}
\def\agt{\mathrel{\hbox{\rlap{\hbox{\lower4pt\hbox{$\sim$}}}\hbox{$>$}}}}

\usepackage{mathtools}
\DeclarePairedDelimiter\abs{\lvert}{\rvert}

\newenvironment{cititemize2}
{\begin{list}{$\bullet$}
        {\setlength{\topsep}{0pt}
         \setlength{\itemsep}{0pt}
         \setlength{\parsep}{0.25\parsep}
         \settowidth{\labelwidth}{$\bullet$}
         \setlength{\leftmargin}{1em}
}
}
{\end{list}}

\def\gta{\ifmmode {\mathbin{\lower 3pt\hbox   
    {$\,\rlap{\raise 5pt\hbox{$\char'076$}}\mathchar"7218\,$}}}
    \else {${\mathbin{\lower 3pt\hbox
    {$\rlap{\raise 5pt\hbox{$\char'076$}}\mathchar"7218\,$}}}
    $}\fi}
\def\lta{\ifmmode {\,\mathbin{\lower 3pt\hbox   
    {$\,\rlap{\raise 5pt\hbox{$\char'074$}}\mathchar"7218\,$}}}
    \else {${\mathbin{\lower 3pt\hbox
    {$\rlap{\raise 5pt\hbox{$\char'074$}}\mathchar"7218\,$}}}
    $}\fi}

\newcommand{\beq}{\begin{equation}}
\newcommand{\eeq}{\end{equation}}
\newcommand{\bea}{\begin{eqnarray}}
\newcommand{\eea}{\end{eqnarray}}

\definecolor{darkperiwinkle}{RGB}{102, 102, 128}

\definecolor{light-gray}{gray}{0.9}

\usepackage[normalem]{ulem}

\begin{document}

\title{Physics-inspired deep learning to characterize the signal manifold of quasi-circular, spinning, non-precessing binary black hole mergers}

\author[NCSA,PNCSA]{Asad Khan}
\author[NCSA,PNCSA,ANCSA]{E. A. Huerta}
\author[NCSA,ECE]{Arnav Das}

\address[NCSA]{National Center for Supercomputing Applications, University of Illinois at Urbana-Champaign, Urbana, Illinois 61801, USA}
\address[PNCSA]{Department of Physics, University of Illinois at Urbana-Champaign, Urbana, Illinois 61801, USA}
\address[ANCSA]{Department of Astronomy, University of Illinois at Urbana-Champaign, Urbana, Illinois 61801, USA}
\address[ECE]{Department of Electrical and Computer Engineering, University of Illinois at Urbana-Champaign, Urbana, Illinois 61801, USA}

\date{\today}

\begin{abstract}
\noindent The spin distribution of binary black hole mergers contains key information concerning the formation channels of these objects, and the astrophysical environments where they form, evolve and coalesce. To quantify the suitability of deep learning to estimate the individual spins, effective spin and mass-ratio of quasi-circular, spinning, non-precessing binary black hole mergers, we introduce a modified version of \texttt{WaveNet} trained with a novel optimization scheme that incorporates general relativistic constraints of the spin properties of astrophysical black holes. The neural network model is trained, validated and tested with 1.5 million \(\ell=|m|=2\) waveforms generated within the regime of validity of \texttt{NRHybSur3dq8}, i.e., mass-ratios \(q\leq8\) and individual black hole spins \(\abs{s}^z_{\{1,\,2\}} \leq 0.8\). To reduce time-to-insight, we deployed a distributed training algorithm at the \texttt{IBM Power9 Hardware-Accelerated Learning} cluster at the National Center for Supercomputing Applications to reduce the training stage from 1 month, using a single \texttt{V100 NVIDIA GPU}, to 12.4 hours using 64 \texttt{V100 NVIDIA GPUs}. We have also fully trained this model using 1536 V100 GPUs (256 nodes) in the Summit supercomputer at Oak Ridge National Laboratory, achieving state-of-the-art accuracy within just 1.2 hours. Using this neural network model, we quantify how accurately we can infer the astrophysical parameters of black hole mergers in the absence of noise. We do this by computing the overlap between waveforms in the testing data set and the corresponding signals whose mass-ratio and individual spins are predicted by our neural network. We find that the convergence of high performance computing and physics-inspired optimization algorithms enable an accurate reconstruction of the mass-ratio and individual spins of binary black hole mergers across the parameter space under consideration. This is a significant step towards an informed utilization of physics-inspired deep learning models to reconstruct the spin distribution of binary black hole mergers in realistic detection scenarios. 
\end{abstract}

\maketitle


\section{Introduction}
\label{sec:intro}
Gravitational wave (GW) observations provide unique insights into the formation channels of compact binary systems. For instance, it is expected that inferring orbital eccentricity in GW observations~\cite{Anton:2014,Huerta:2009,antoni:2018A,Anto:2015arXiv,Adam:2018prd,Huerta:2019oxn,habhu:2019,johnson:2017,huerta:2018PhRvD,Huerta:2017a,hinder:2017a,cao:2017,Hinderer:2017} may provide the most conclusive evidence for the existence of compact binary systems in dense stellar environments~\cite{galcen:2018,Sippel:2013,Strader:2012,ssm:2018,lisa:2018b,sam:2018MNRAS1548S,Huerta:2015a,sam:2019MNRAS30S,Huerta:2014,samdor:2018MNRAS5445S,samdorII:2018MNRAS4775D,rocarl:2018PhRvDR,kremerjoh:2018aK,lopez:2018L,hoang:2017APJ,lisa:2018a,antonras:2016ApJ7A,Huerta:2013a,sam:2017ApJ,Leigh:2018MNRAS,ssm:2017}. For instance, in the case of binary black hole (BBH) mergers, it is assumed that the spin distribution of BBHs formed in dense stellar environments may be distributed isotropically, whereas BBHs formed through massive stellar evolution in isolation may have spin distributions that are aligned with the binary's orbital angular momentum~\cite{Farr:2017gtv,Fernandez:2019kyb}.

As the number of GW observations of BBH mergers continues to grow in years to come~\cite{o1o2catalog,Abbott_2020_ns,LIGOScientific:2020stg}, it will be possible to infer the astrophysical properties of these sources and elucidate their formation history. 

The goal of this article is to explore how deep learning handles parameter space degeneracies in the signal manifold of quasi-circular, spinning, non-precessing BBH mergers; and to quantify how accurately deep learning may constrain the individual spins, effective spin and mass-ratio of these GW sources in the absence of noise. We then go on to discuss the computational grand challenges that naturally arise when one tries to address these problems: (1) the parameter space that needs to be sampled is very large, requiring the use of TB-size waveform datasets, and thereby demanding the development of novel distributed algorithms to use many GPUs to fully train deep learning algorithms in a reasonable amount of time; and (2) the need to incorporate domain knowledge into the optimization of deep learning algorithms to accelerate their convergence and to ensure that their predictions are physically consistent.

In connection to the first challenge mentioned above, we introduce distributed training algorithms that reduces the training stage of the neural network model used in this study from one month using a single V100 GPU to: (i) 12.4 hours using 64 V100 GPUs at the Hardware Accelerated Learning (HAL) cluster at the National Center for Supercomputing Applications (NCSA); and (ii) within 1.2 hours using 1536 V100 GPUs at the Summit supercomputer at Oak Ridge National Laboratory. These results establish a record in the number of GPUs used to train these types of physics-inspired deep learning models.

Regarding the second challenge, we have found that naive methods to train deep learning architectures lead to rather sup-optimal results. However, we show that when we use physics-inspired optimization algorithms, which incorporate general relativistic constraints of the spin of BBHs, we are able to accurately recover the individual spins and mass-ratio of BBH signals across the mass-ratio under consideration. This analysis provides benchmarks for the performance of deep learning models when applied to the reconstruction of these parameters in the absence of noise, and provides a baseline of accuracy when real noise from GW observations is taken into consideration. That study will be presented shortly in a follow up paper.

To describe the signal manifold of quasi-circular, spinning, non-precessing BBH mergers, we use a catalog of over one million time-series waveforms that are parameterized in natural units, i.e., we use numerical relativity (NR) type waveforms, such that the GWs that describe BBH mergers may be fully described in terms of mass-ratio, \(q\), and the individual spins of the binary components, \( s^z_i\) with \(i=\{1,\,2\}\). The neural network will be trained to cover this 3-D signal manifold \((q,\, s^z_1,\, s^z_2)\), with the aim of accurately inferring these three parameters when unlabelled waveforms are fed into the neural network. 

This work builds upon an emergent field of deep learning research which has thus far provided new methodologies to do detection and point-parameter estimation for GW sources in the context of simulated noise~\cite{geodf:2017a}, and real advanced LIGO noise~\cite{geodf:2017b,geodf:2017c}; detection-only methods in the context of simulated noise~\cite{gabb:2018,fan:2019SCPMA}, real noise~\cite{PhysRevD.100.063015}; denoising of GW signals~\cite{Shen:ICASP2019,Wei:2019zlc,Torres:2020}, among others (see~\cite{huerta:2019NatRP} and references therein). In the context of other recent deep learning studies

\begin{itemize}
    \item Our work differs from~\cite{yamamoto2020use} in that they focus on the physics of waveform signals during the ringdown phase, whereas we are interested in the effective spin or individual spins of BBH mergers through the waveform evolution.
    \item The authors in~\cite{gabbard2019bayesian} use conditional variational autoencoders to produce fast posterior probability estimates for \textit{non-spinning} BBH mergers embedded in simulated noise. Furthermore,  the authors in~\cite{green2020gravitationalwave} introduce the use of auto-regressive normalizing flows for rapid inference. They use a small data set of spinning BBH mergers, and assume signals with signal-to-noise ratios \(\gtrsim 20\), to produce marginalized posterior distributions for a number of parameters, including the individual spins and effective spin of non-precessing BBH mergers in the context of simulated noise. In contrast, in this paper we focus on the development of physics-inspired neural network models that, in the absence of noise, provide insights about how deep learning handles parameter space degeneracies and the measurement of individual spins, effective spin and mass-ratio of spinning, non-precessing BBH mergers. In brief, herein we develop a computational framework to minimize time-to-insight by combining extreme scale computing and physics-inspired models. This approach will be exploited in a future study to carefully assess how simulated and real noise bias the results we present in this article.
\end{itemize}

\noindent The previous list shows that it is timely and relevant to start developing deep learning models that encapsulate as much physics as possible, which may only be accomplished by using larger training datasets and developing physics-inspired models and optimization schemes. In turn, this demands the development of distributed training algorithms in high performance computing platforms to reduce time-to-solution. This article is a significant step in that direction.

This study is organized as follows. Section~\ref{sec:method} describes the data sets used to train, validate and test our neural network model. We also describe therein the deep learning model used, and show how to build a physics-inspired optimization algorithm that significantly improves the predictive accuracy of our neural network. Our findings are presented and discussed in Section~\ref{sec:res}. We summarize this work and outline future directions of work in Section~\ref{sec:end}. 

\section{Methods}
\label{sec:method}

In this section we describe the data sets used to train, validate and test our neural network model. We also describe the neural network architecture used for this work, and the construction of a physics-inspired optimization algorithm that significantly improves the ability of our neural network to accurately infer the astrophysical properties of BBH mergers from the time-series NR waveforms that describe them. 

\subsection{Data Curation}
\label{sec:datcuration}

We generate our training, validation and test sets using \texttt{NRHybSur3dq8} \cite{PhysRevD.99.064045}, a surrogate model for hybridized non-precessing NR waveforms. While this surrogate model may be extrapolated up to mass-ratio \( q = 10\) and \( s^z_i = 0.998\), with \(i=\{1,\,2\}\), it has only been trained with 104 NR waveforms in the parameter range \(q \leq 8\) and \(s^z_i \leq 0.8\), and hence we restrict our data sets to the same parameter span. Furthermore, we consider the \(\ell=\abs{m}=2\) mode to train, validate and test our neural network model. To be consistent with the data used to train, validate and test our neural network, throughout this paper we use a geometric unit system in which \(G=c=1\).

\begin{figure}[h!]
\centerline{
\includegraphics[width=\linewidth]{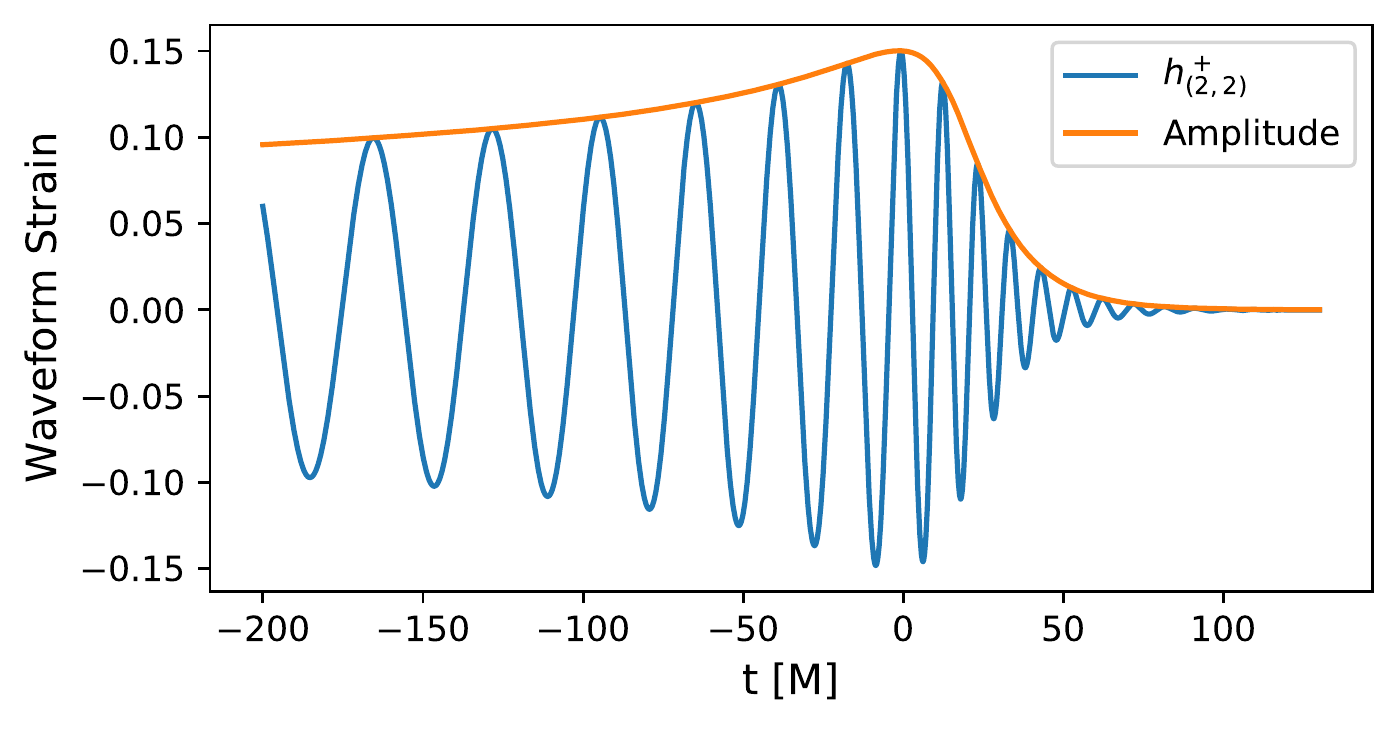}
} 
\caption{Sample waveform for a binary black hole system with mass-ratio $q=7.9$, and whose binary components have spins  $s^z_1 = s^z_2 = 0.8$, respectively. All waveforms used in this analysis cover the time range \(t\in[-10,000\,\textrm{M}, 130\,\textrm{M}]\), and are sampled with a time step \(\Delta t = 0.1\,\textrm{M}\).}
\label{fig:sample_wf}
\end{figure}

The signals produced for this analysis are such that the amplitude peak occurs at \(t=0M\), covering a time span \(t\in[-10,000\,\textrm{M}, 130\,\textrm{M}]\) with a time step \(\Delta t = 0.1\,\textrm{M}\). A sample waveform is shown in Fig.~\ref{fig:sample_wf}.

\begin{figure}[h!]
\centerline{
\includegraphics[width=\linewidth]{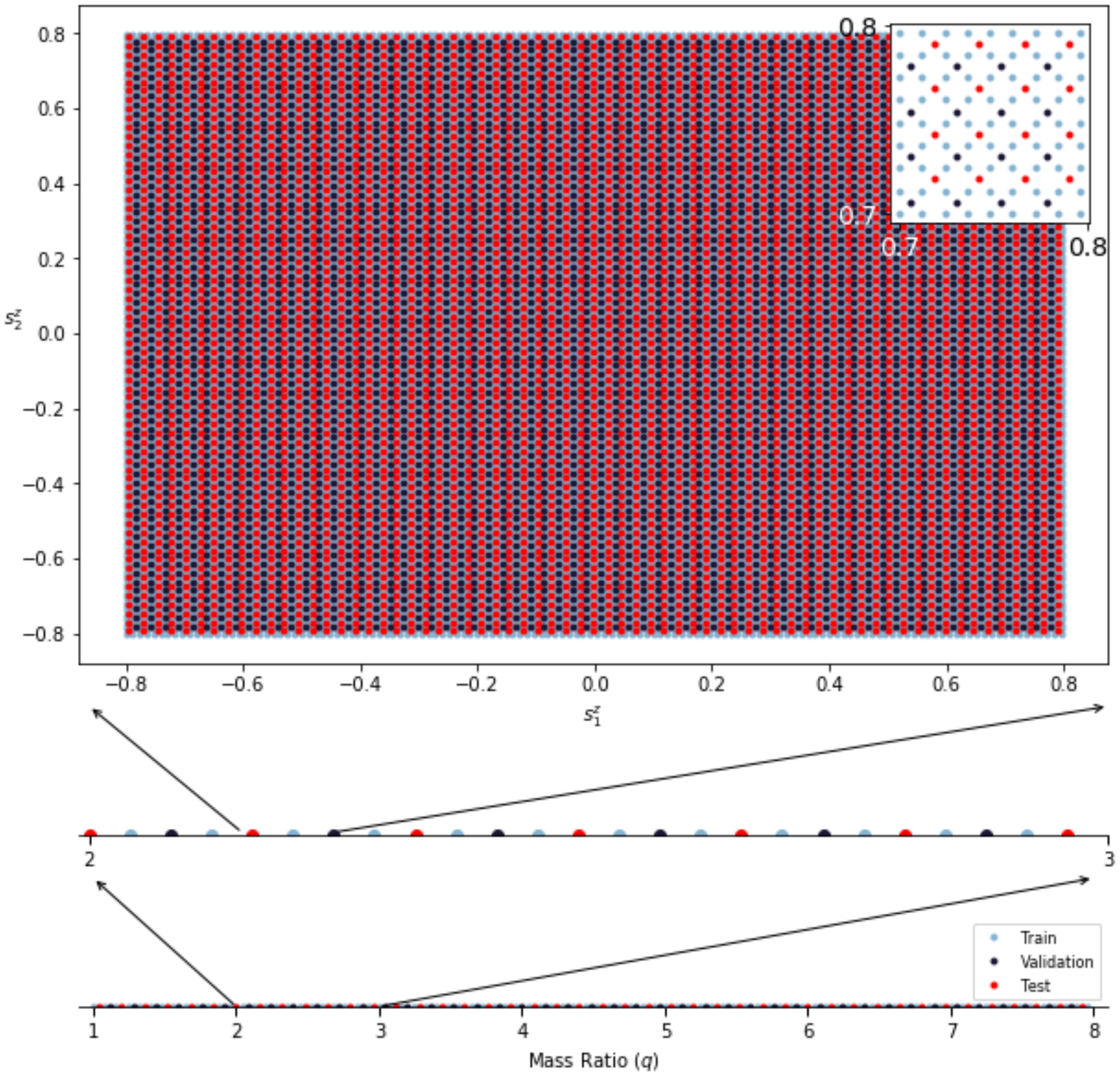}
} 
\caption{Sampling of the signal manifold $q\in[1,\,8]$, $s^z_{\{1,\,2\}}\in[-0.8,\,0.8]$ to construct the training (light blue dots), validation (dark blue dots) and testing (red dots) data sets.}
\label{fig:data_distribution}
\end{figure}

The training set is generated by sampling the mass-ratio $q\in[1,8]$ in steps of \(\Delta q = 0.08\); and the individual spins  $s^z_i\in[-0.8, 0.8]$ in steps of \(\Delta s^z_i = 0.012\). This is equivalent to $\sim1.5$ million waveforms. The validation and test sets are generated by alternately sampling the intermediate values, i.e. by sampling $q$ and $s^z_i$ in steps of $0.16$ and $0.024$ to lie between training set values, for a total of $\sim 190,000$ waveforms each respectively. The distributions of parameters for training, validation and test sets is summarized in Fig~\ref{fig:data_distribution}. 
The entire data set is $\sim1.5$TB in size, and we make use of \texttt{mpi4py} \cite{DALCIN20111124, DALCIN2008655, DALCIN20051108} to parallelize the data generation using the Campus Cluster at the University of Illinois at Urbana-Champaign~\cite{campuscluster}.

\subsection{Neural Network Model Architecture and Loss Function}
\label{sec:DeepLearning}

The neural network architecture consists of two fundamental components, a shared base/root consisting of layers slightly modified from the \texttt{WaveNet} \cite{oord2016wavenet} architecture, and two branches consisting of fully connected layers that take in features extracted from the root to predict the mass ratio and the individual spins of the binary components, respectively.

\texttt{WaveNet} is a probabilistic and auto-regressive model originally released by \texttt{DeepMind} for generating raw audio waveforms, with predictive distribution for each audio sample conditioned on all previous ones. Trained on data with tens of thousands of samples per second of audio, it exhibited not only state of the art performance on text-to-speech, but also showed promising results in music generation and as a discriminative model for phoneme/speech recognition. Additionally \texttt{WaveNet} inspired architectures have also been demonstrated to successfully denoise recent GW observations~\cite{Wei:2019zlc}. Inspired by such recent successes, as well as key architectural features (which we describe in more detail below) suited to processing wideband raw waveforms, we modify the \texttt{WaveNet} architecture and test it as a discriminative model to predict key parameters of GWs from quasi-circular, spinning, non-precessing binary black hole mergers.


The key architectural components of a \texttt{WaveNet} are dilated causal convolutions, gated activation units, and the usage of residual and skip connections, which we describe below.

\subsubsection{Root Layers}

\paragraph{Dilated Causal Convolutions}

The original \texttt{WaveNet} was used as a generative model to predict audio sample $x_t$ conditioned on all previous timesteps, where the joint probability of the waveform $\boldsymbol{x} = \{x_1, x_2, ..., x_T\}$ is factorized as a product of conditional probabilities $p(\boldsymbol{x}) = \prod_{t=1}^T p(x_t | x_1, ..., x_{t-1})$. Traditionally recurrent neural networks (RNNs) and Long short-term memory (LSTM) models have been used to model conditional probabilities of sequences, but they are typically considerably slower than convolutional networks and also suffer from vanishing gradient problem for very long sequences. \texttt{WaveNet} addressed the aforementioned issues using causal convolutions, which are implemented by shifting the output of a normal convolution by a few timesteps, and ensuring that the prediction made by the model at time step $t$ does not depend on the future timesteps $x_{t+1}, ..., x_{T}$. Since ours is a discriminative model, we turn off the causality of the \texttt{WaveNet} architecture, so that our model can simultaneously process all time steps of the waveform.

\begin{figure}
    \centering
    \includegraphics{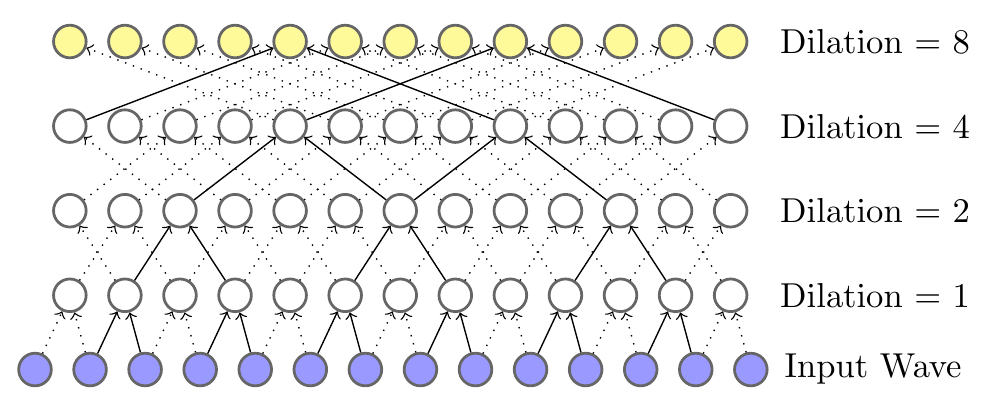}
    \caption{Stack of Dilated Convolutional Layers. Zero padding on either side is not shown.}
    \label{fig:DilatedConv}
\end{figure}

A dilated convolution is an effective way to increase the receptive field of a convolutional network by orders of magnitude with only a small computational overhead. This is achieved by applying a convolutional filter over the input sequence by skipping input values with a certain step size (called the `dilation'). Figure~\ref{fig:DilatedConv} shows dilated convolutions for dilations of 1, 2, 4, and 8 respectively. Similar to the original \texttt{WaveNet} architecture, we double the dilation for every layer up to a limit and then repeat the same stack of layers, to efficiently capture long range structure of the waveform.

\paragraph{Gated Activation Units}

A potential advantage of using LSTMS is that they have multiplicative units (memory gates) that may help with modeling complex inter-dependencies between time steps. In the original \texttt{WaveNet} architecture, this is amended by replacing the rectified linear units (ReLU) between convolutions with the following gated activation unit:

\begin{equation}
    \boldsymbol{z} = \tanh (W_{f,k} * \boldsymbol{x}) \odot \sigma (W_{g,k} * \boldsymbol{x})\,,
    \label{eq:activation_unit}
\end{equation}

\noindent where $*$ is a convolution operator, $\odot$ is an element-wise multiplication operator, $\sigma (.)$ is the sigmoid function, $k$ is the layer index, and $W_f$ and $W_g$ denote filter and gate convolutions, respectively. We keep the same gated activation Units in our network.

\paragraph{Residual and Skip Connections}

Similar to the original \texttt{WaveNet} architecture, both residual and skip connections are used for fast convergence by resolving vanishing gradients when training deep neural networks. Fig~\ref{fig:architecture} shows a residual block of the model, which is stacked many times in the network. More in depth discussion of the structure of \texttt{WaveNet} may be found in the original paper~\cite{WaveNet}.

\subsubsection{Leaf Layers}

\begin{figure}
    \centering
    \includegraphics[width=\linewidth]{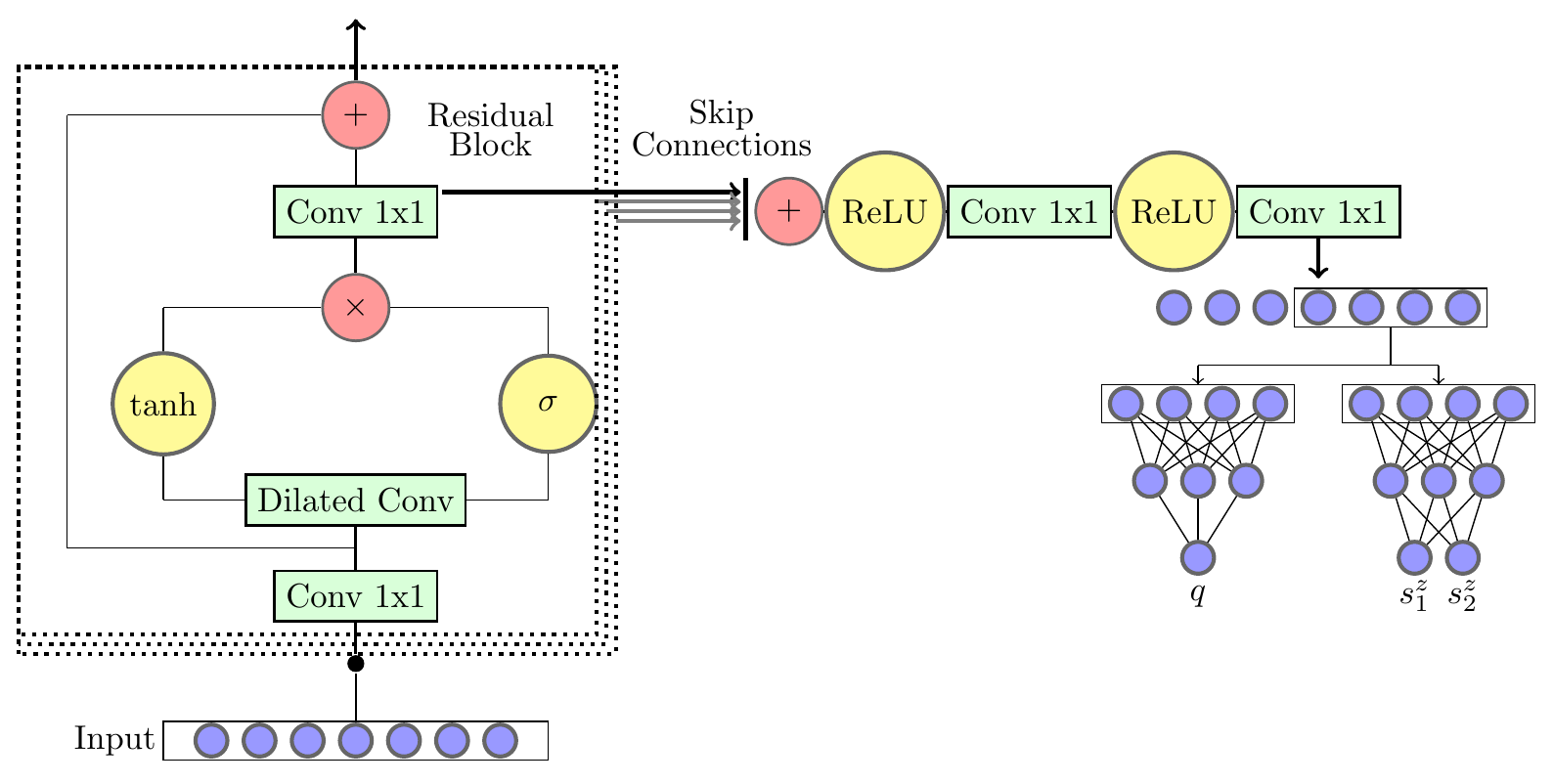}
    \caption{Network Architecture. Residual Blocks of Wavenet (left) and leaf layers (right).}
    \label{fig:architecture}
\end{figure}

At the end of root layer the model outputs a sequence with the same time dimensionality as the input sequence. Since the input sequence has $101,300$ time-steps, we pick the last $10,000$ time-steps from the output sequence, flatten and feed into two leafs of fully connected layers predicting the mass ratio \(q\) and spins \(s^z_i\) respectively, as shown in Fig~\ref{fig:architecture}. Hence the shared root can be thought of as a feature extractor, and the leafs as sub-networks specializing to predict different parameters from the extracted features. The motivation behind this root/leaf structure is two-fold; firstly the mass ratio \(q \in [1, 8]\) and the spins \(s^z_i \in [-0.8, 0.8]\) have different numerical ranges, and secondly we performed experiments without a leaf structure and empirically got sub-optimal results.

\subsubsection{Loss Function}

We performed several experiments to find an optimal way to constrain \(s^z_i\) predictions from the second leaf sub-network. In the first iteration, we predicted \(s^z_i\) directly using mean-squared error as the loss function. This resulted in reasonable predictions for \(s^z_1\), but \(s^z_2\) predictions could not be constrained with continued training. Consequently, we explored the use of effective one-body general relativistic dynamics of two spinning black holes as delineated in \cite{Damour:2001tu}. Specifically, we focus on the derivation of a spin-dependent effective one-body Hamiltonian for small and moderate spins as a \(\nu\)-deformation of a Kerr metric of mass \(M \equiv m_1 + m_2\) and effective spin 

\begin{equation}
    S_{\textrm{eff}} = \sigma_1 s^z_1 + \sigma_2 s^z_2\,,
    \label{eq:eff_spin}
\end{equation}

\noindent where \(\sigma_1 \equiv 1 + \frac{3}{4q}\) and \(\sigma_2 \equiv 1 + \frac{3q}{4}\). In addition, we also consider the effective spin parameter~\cite{Farr:2017gtv,TheLIGOScientific:2016wfe} 

\begin{equation}
    \sigma_{\textrm{eff}} = \frac{m_1s^z_1 + m_2 s^z_2}{m_1 + m_2} = \frac{q s^z_1 + s^z_2}{1 + q}\,.
    \label{eq:alt_eff_spin}
\end{equation}

\noindent Predicting \( S_{\textrm{eff}}\) and \(\sigma_{\textrm{eff}}\), and using our prediction for \(q\) to solve Equations~\eqref{eq:eff_spin} and~\eqref{eq:alt_eff_spin}, we were able to tightly constrain \(s^z_i\) predictions.

\section{Results}
\label{sec:res}

This section presents our main findings in the following format: we first discuss specific challenges to be addressed, namely, the parameter space degeneracy of the signal manifold under consideration and the need to use a large training data set to densely sample this 3-D signal manifold. We then present qualitative results of the performance of our neural network model, and finalize with a quantitative set of results that provide a thorough description of the realm of applicability of our neural network model. 

\subsection{Parameter space degeneracy and convergence of high performance computing with deep learning}

\begin{figure*}[h!]
\centerline{
\includegraphics[width=1.0\linewidth]{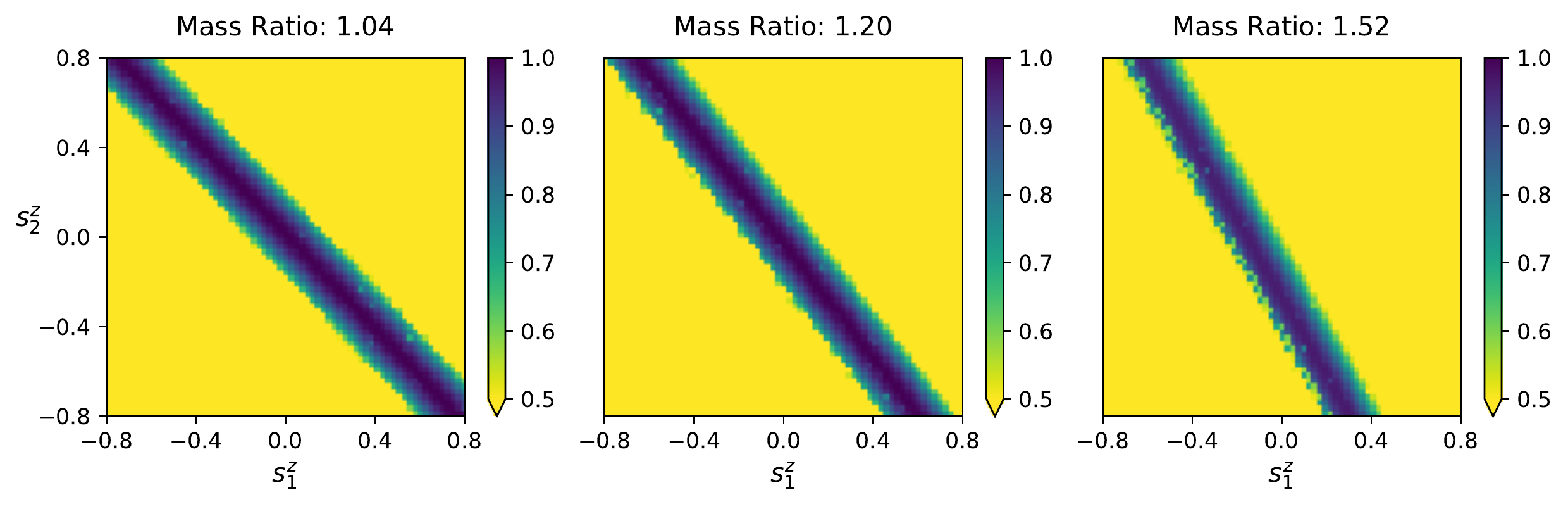}
} 
\centerline{
\includegraphics[width=1.0\linewidth]{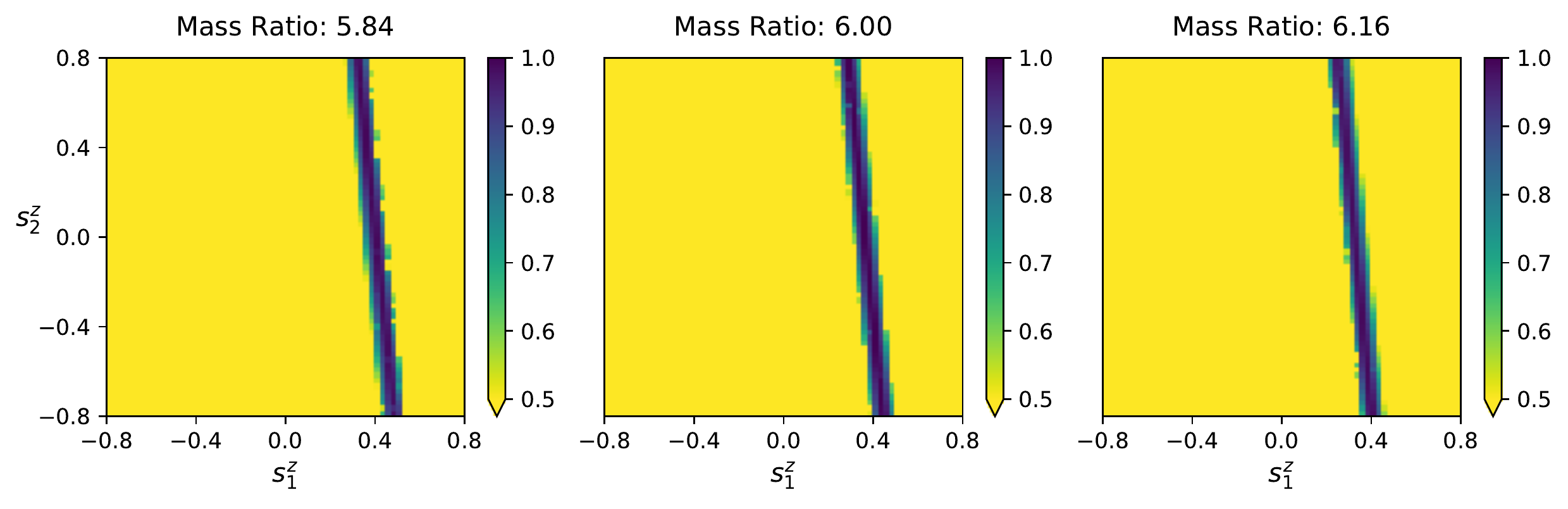}
} 
\centerline{
\includegraphics[width=1.0\linewidth]{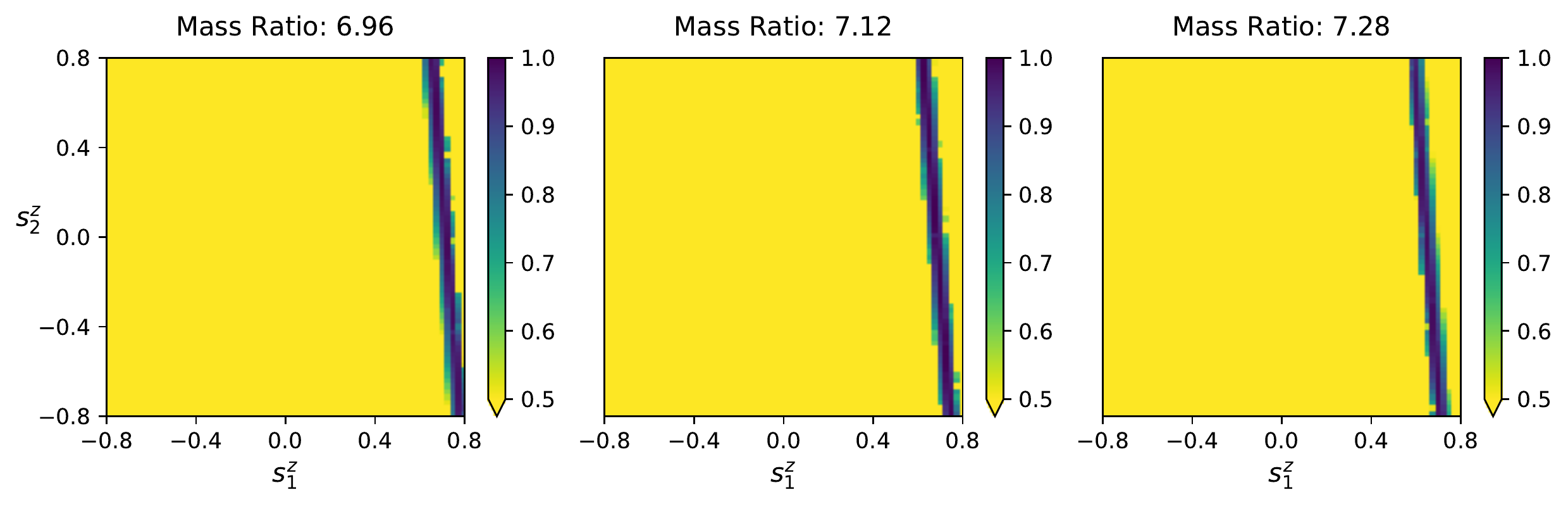}
} 
\centerline{
\includegraphics[width=1.0\linewidth]{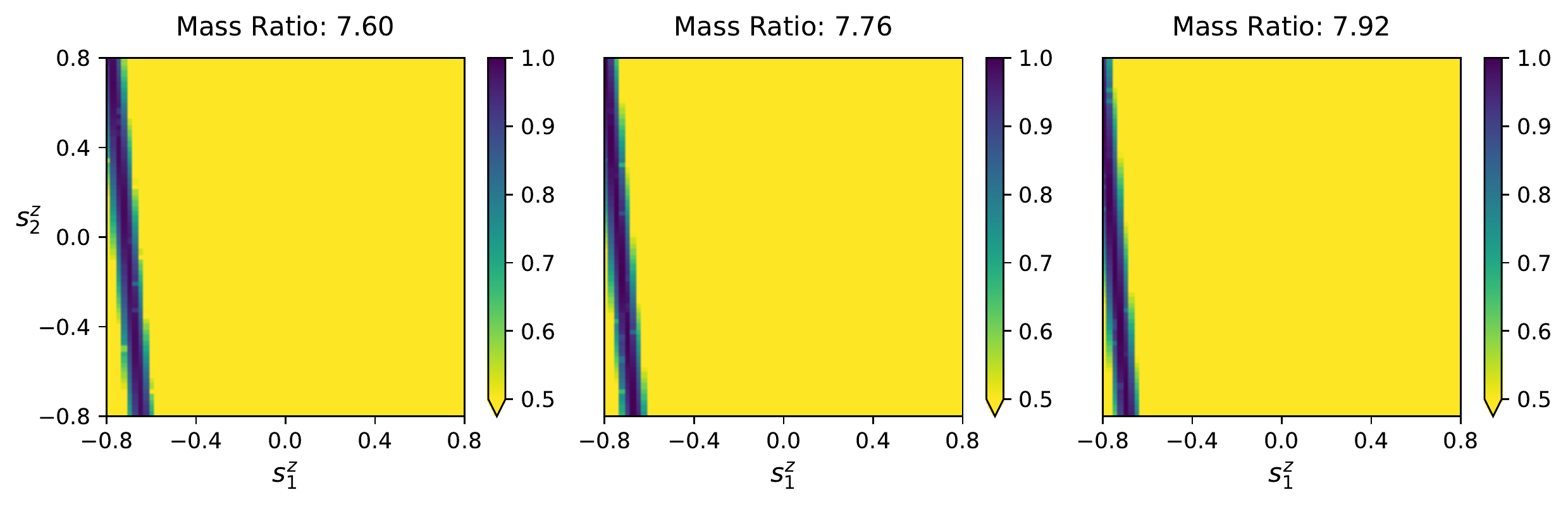}
} 
\caption{From top to bottom, each row presents regions of parameter space degeneracy for a random sample of binary black hole mergers with parameters: (i) $q=1.0, s^z_1=0.8, s^z_2=-0.8$; (ii)  $q=6.00, s^z_1=0.40, s^z_2=-0.40$; (iii) $q=7.12, s^z_1=0.70, s^z_2=-0.30$; and (iv) $q=7.92, s^z_1=-0.70, s^z_2=-0.70$. The color bar indicates the overlap between each of the four aforementioned binary black hole mergers and the three corresponding mass-ratio slices shown on each row.}
\label{fig:degeneracy}
\end{figure*}

\noindent Inferring key parameters from the signal manifold of quasi-circular, spinning, non-precessing, BBH mergers presents a number of complications given that different time-series NR waveforms, say \(h(t)\) and \(s(t)\), that describe different astrophysical systems are remarkably similar. To illustrate this property we have selected a few astrophysical systems, and then computed the overlap, \({\cal{O}}  (h,\,s)\), between them and 2-D slices of the signal manifold, determined by the BBH mass-ratio and \(s^z_{\{1,\,2\}}\), as shown in Figure~\ref{fig:degeneracy}, using the relation

\begin{equation}
\label{over}
{\cal{O}}  (h,\,s)= \underset{ t_c\, \phi_c}{\mathrm{max}}\left(\hat{h}|\hat{s}_{[t_c,\,  \phi_c]}\right)\,,\quad{\rm with}\quad \hat{h}=h\,\left(h | h\right)^{-1/2}\,,
\end{equation}

\begin{figure*}[h!]
\centerline{
\includegraphics[width=0.5\linewidth]{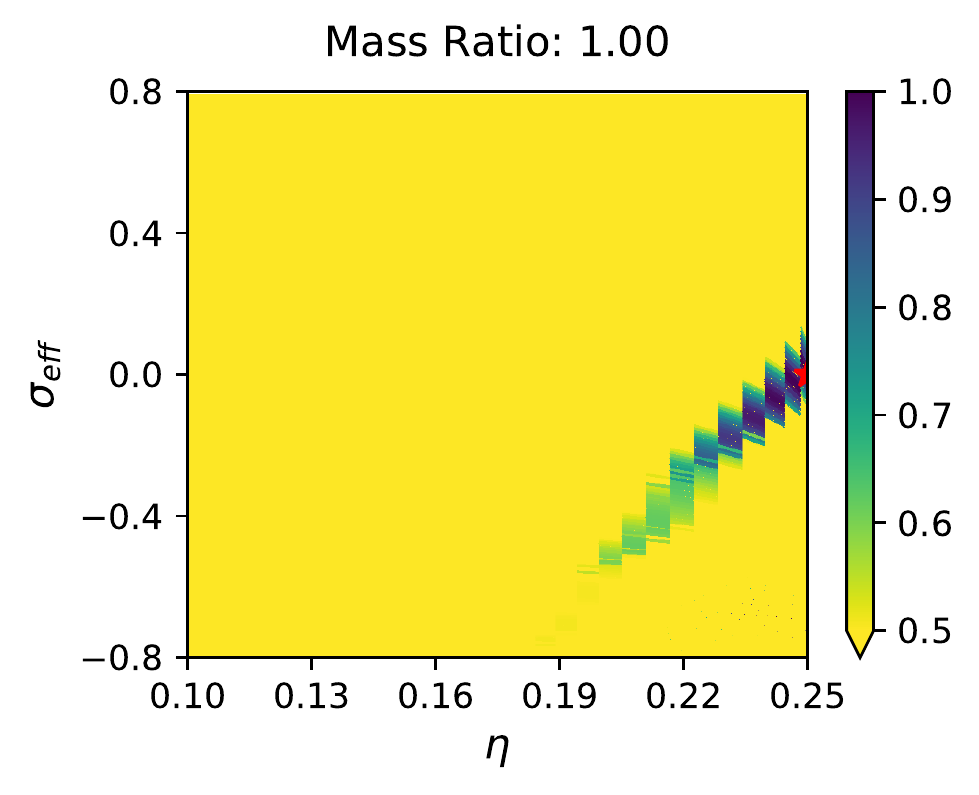}
\includegraphics[width=0.5\linewidth]{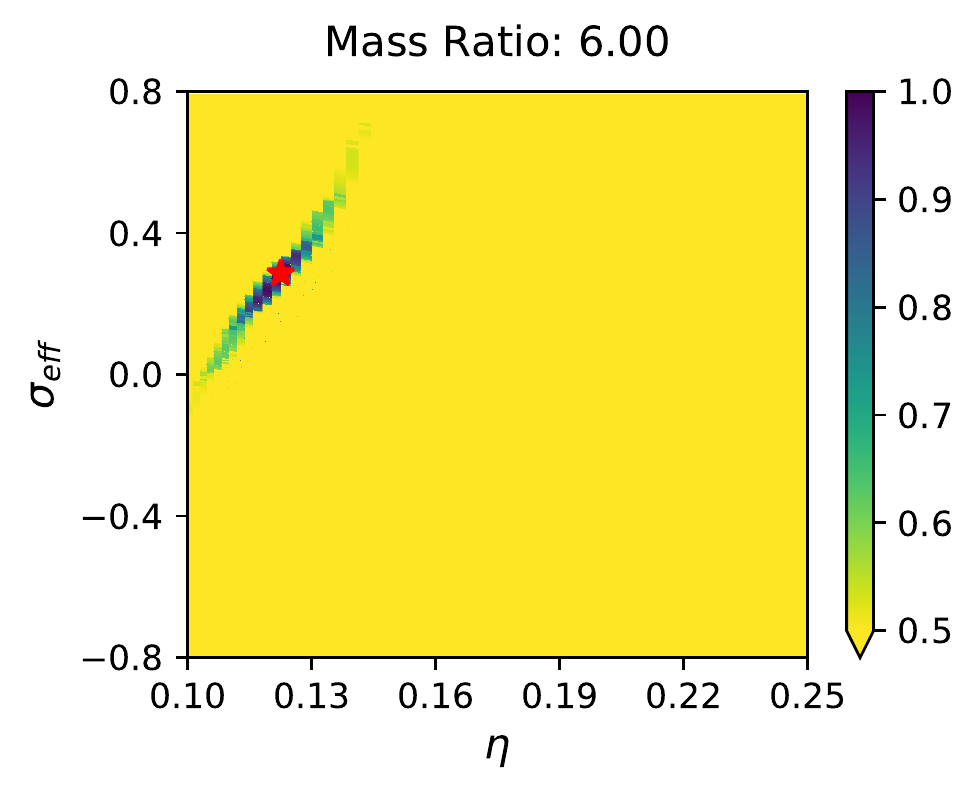}
}
\centerline{
\includegraphics[width=0.5\linewidth]{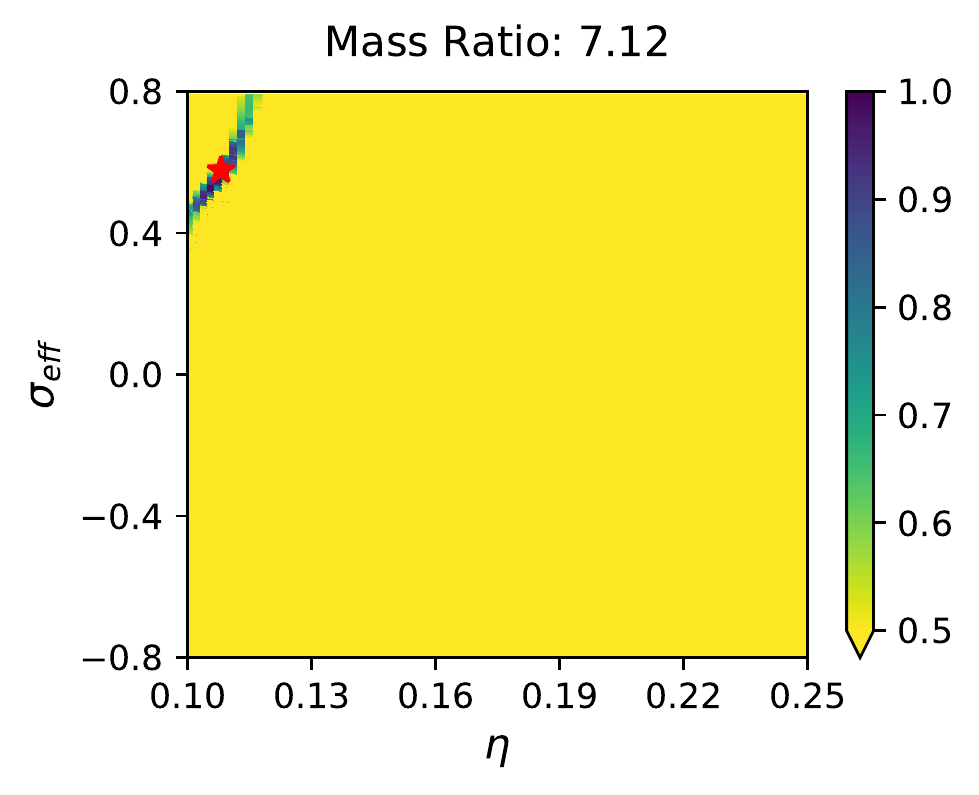}
\includegraphics[width=0.5\linewidth]{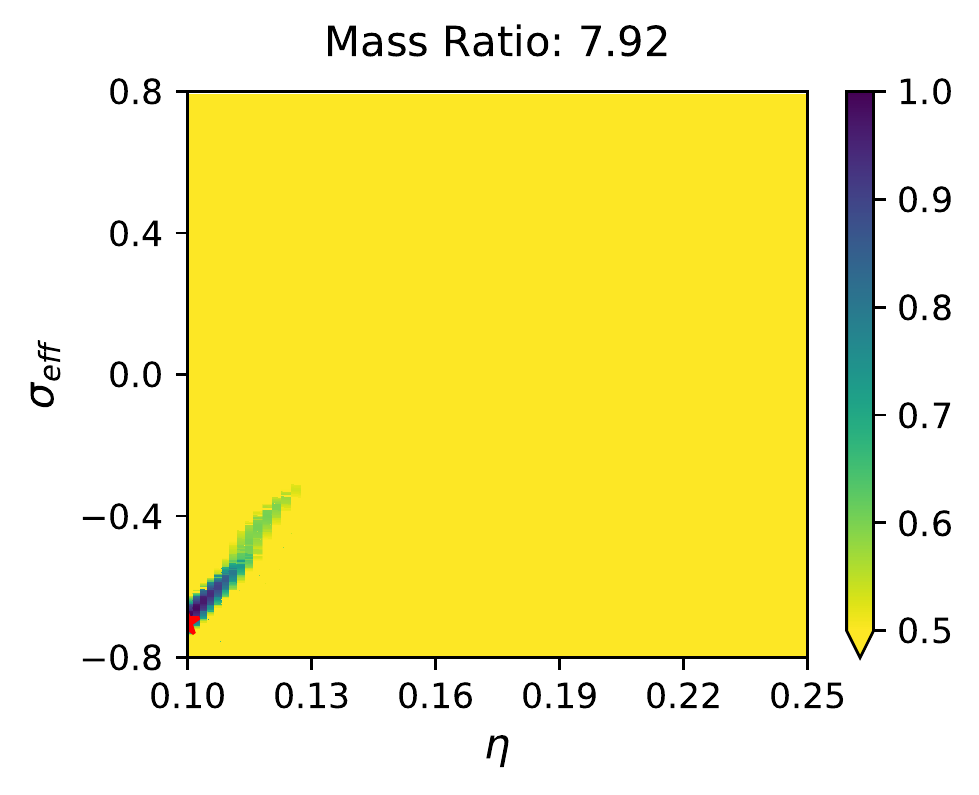}
} 
\caption{As Figure~\ref{fig:degeneracy}, but now presenting results in terms of symmetric mass ratio, $\eta$, and effective spin, $\sigma_{\textrm{eff}}$. True parameters are indicated by red stars. The color bar indicates the waveform overlap between true parameters (red star) and the rest of the parameter space.}
\label{fig:degeneracy_leading_order_correction}
\end{figure*}

\noindent where \(\hat{s}_{[t_c,\,  \phi_c]}\)  indicates that the normalized waveform \(\hat{s}\)  has been time- and phase-shifted. Using this metric, Figure~\ref{fig:degeneracy} shows that the overlap between a given signal and other BBH systems with remarkably different spin combinations are effectively indistinguishable. Other observations we obtain from these results is that inferring the individual spins of near-equal mass-ratio systems is very hard given the intrinsic symmetry of the system, i.e., the binary components may be interchanged \(m_1\Longleftrightarrow m_2\), as shown in the top panels of Figure~\ref{fig:degeneracy}. Indeed, near equal BBH systems present the largest regions of degeneracy across the signal manifold. We also observe that systems with small to moderate spins present large regions of degeneracy, as shown in the second and third row of panels in Figure~\ref{fig:degeneracy}. These features are also present in BBHs whose binary components have large spin values, as shown in the bottom panels of Figure~\ref{fig:degeneracy}. Following~\cite{Purrer:2015nkh}, we recast the results in Figure~\ref{fig:degeneracy} in terms of the symmetric mass-ratio, \(\eta\), and the effective spin, \(\sigma_{\textrm{eff}}\), where 
\begin{equation}
    \eta = \frac{m_1 m_2}{(m_1+m_2)^2}
\end{equation} These results provide a holistic perspective on spin and mass-ratio degeneracy across the parameter space under consideration as shown in Figure~\ref{fig:degeneracy_leading_order_correction}. In summary, inferring the individual spins of BBH mergers is a rather challenging endeavor. 


\begin{figure}[h!]
\centerline{
\includegraphics[width=0.5\linewidth]{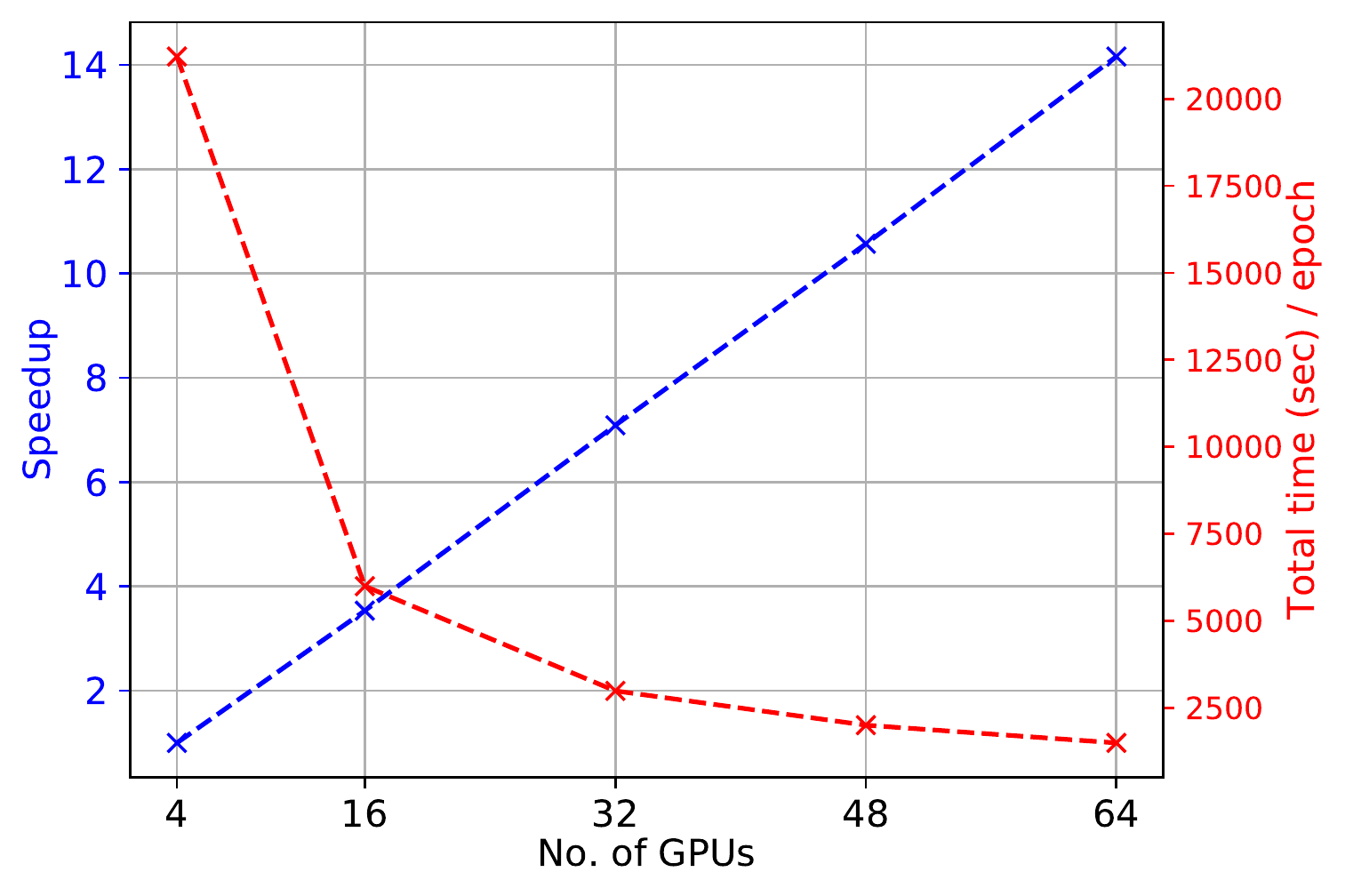}
\includegraphics[width=0.5\linewidth]{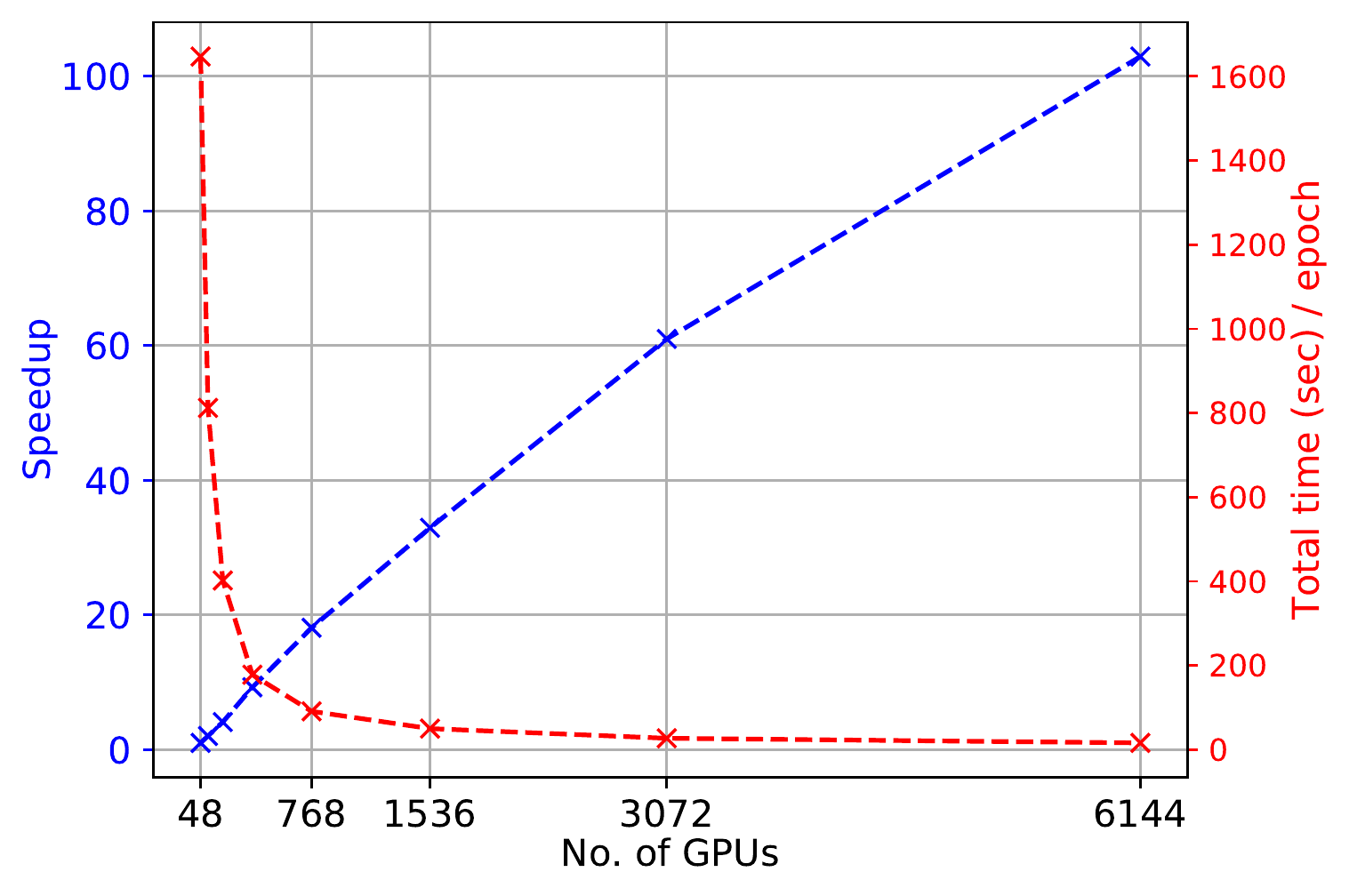}
} 
\caption{ Left: Speed up in training obtained by designing and deploying a distributed training scheme at the Hardware-Accelerated Learning (HAL) deep learning cluster at the National Center for Supercomputing Applications. This approach reduces the training stage from one month (using a single V100 GPU) to 12.4 hours by scaling the training to the entire cluster, which consists of 64 NVIDIA V100 GPUs distributed evenly across 16 nodes, and connected by NVLink 2.0 inside the nodes and EDR InfiniBand across the nodes~\cite{halcluster}. Right: Speed up obtained upon deploying and tuning our distributed training scheme on the Summit supercomputer at Oak Ridge National Laboratory.}
\label{fig:HAL_Scaling}
\end{figure}

To try to address these challenges, we begin by densely sampling the signal manifold under consideration following the methodology described in Section~\ref{sec:datcuration}, which leads to the construction of a data set that includes over \(1.5\times10^6\) waveforms. Training the model described in Section~\ref{sec:DeepLearning} with this large data set using a single V100 GPU would take about 30 days to ensure that the model converges and attains optimal performance, which is quantified by computing the overlap between ground-truth signals and signals whose parameters are predicted by the neural network. In order to minimize time-to-insight and to test multiple, physics-inspired optimization algorithms, we had to design and deploy a data-parallel distributed training scheme on the Hardware-Accelerated Learning (HAL) deep learning cluster at the National Center for Supercomputing Applications~\cite{halcluster,huerta_scaling}. This cluster has 64 NVIDIA V100 GPUs distributed evenly across 16 nodes, and connected by NVLink 2.0 inside the nodes and EDR InfiniBand across the nodes. As shown in the left panel of Figure~\ref{fig:HAL_Scaling}, this approach reduces the training stage to only 12.4 hours. Additionally, we also scaled the data-parallel distributed training strategy up to 6144 NVIDIA V100 GPUs at $80\%$ efficiency on the Summit supercomputer at Oak Ridge National Laboratory, as shown in right panel of Figure~\ref{fig:HAL_Scaling}. In data-parallel distribution scheme, the neural network model is replicated on each individual GPU and each replication is fed non-overlapping batches of training data in parallel. After each batch, the GPUs communicate to synchronize gradients and update model weights. Because this scheme involves a linear increase in global batch size with the number of GPUs, it has been observed that such scaling leads to a degradation in convergence and generalization of the model \cite{keskar2016largebatch}. To address this issue, we employed the layer-wise adaptive large batch optimization technique (LAMB)~\cite{you2019large}, and successfully trained the model without degradation in convergence using 1536 NVIDIA V100 GPUs within 1.2 hours. We have made the trained model~\cite{dlhubmodel2} and testing dataset publicly available at Data and Learning Hub for Science (DLHub)~\cite{blaiszik_foster_2019, dlhub} hosted at Argonne National Laboratory.

\subsection{Qualitative Analysis}

\begin{figure}[h!]
\centerline{
\includegraphics[width=1.0\linewidth]{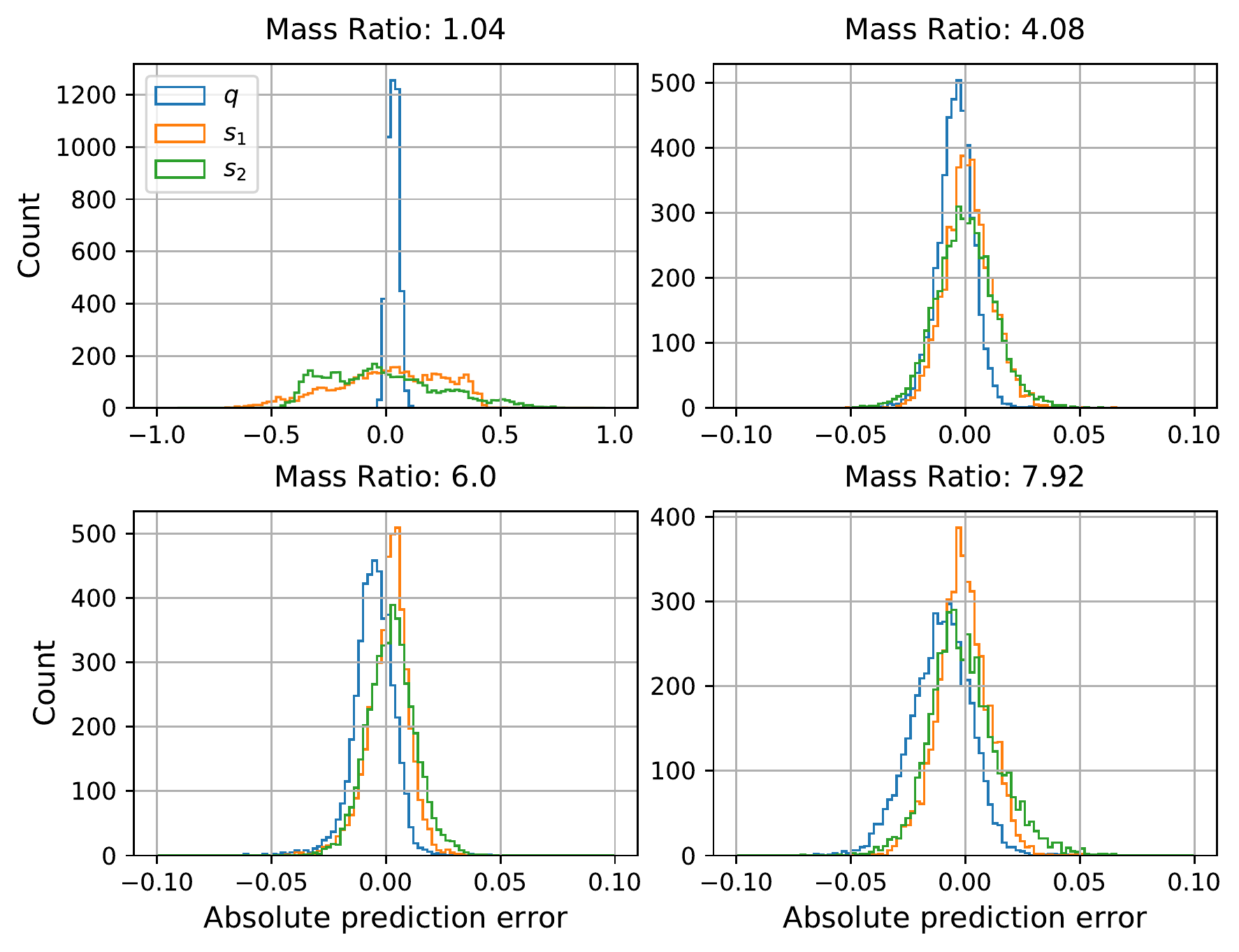}
} 
\centerline{
\includegraphics[width=0.5\linewidth]{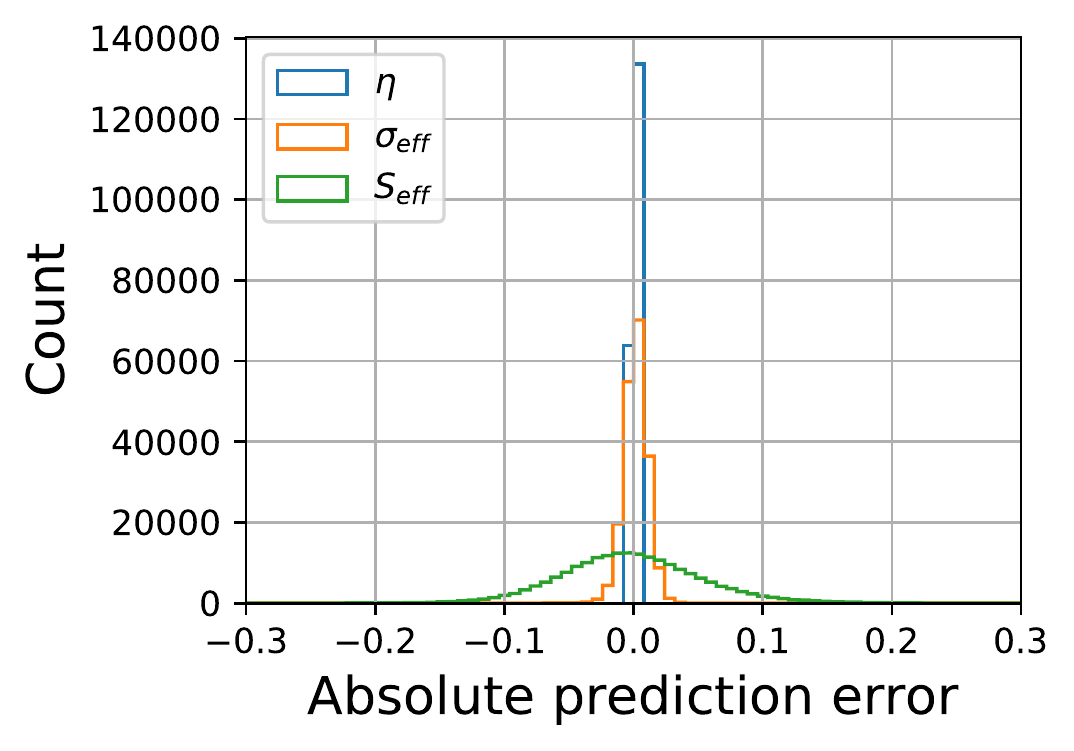}
}
\caption{Top and middle panels: Absolute prediction errors in the recovery of \((q,\,s_1^z,\,s_2^z)\) for a sample of mass ratios. Bottom panel: Absolute prediction errors in the recovery of symmetric mass-ratio,,effective spin and Hamiltonian effective-spin, \((\eta,\, \sigma_{\textrm{eff}},\, S_{\textrm{eff}})\), across the parameter space under consideration.}
\label{fig:abs_err_hist}
\end{figure}

\noindent The first assessment we have conducted to examine the performance of our fully trained model is presented in Figure~\ref{fig:abs_err_hist}. Therein we show the absolute errors in the estimation of the mass-ratio and individual spins for several fixed, mass-ratio slices of the signal manifold, where the absolute error $\Delta{Q_i}$ for any quantity $Q_i$ is defined as: \begin{equation} \Delta{Q_i} = Q_i^{prediction} - Q_i^{true} \end{equation}The top left panel shows again the degeneracy expected for near equal mass-ratio systems. Note that while we can obtain a good estimate of the mass-ratio of these systems, it is hard to infer the individual spins of the binary components. Once this degeneracy is broken, i.e., we consider \(q> 1\) BBH mergers, it is now possible to infer the individual spins and the mass-ratio of the BBH mergers with the accuracy shown in the top-right and bottom-left panels in Figure~\ref{fig:abs_err_hist}. As expected, as the mass-ratio of the BBH increases, it becomes increasingly difficult to infer the spin of the secondary---see bottom right panel of Figure~\ref{fig:abs_err_hist}. This is expected, since for asymmetric mass-ratio BBH mergers, the effect of the secondary in the dynamics of the binary system becomes less dominant, i.e., for a fixed mass-ratio and spin of the primary, it is difficult to tell apart signals when we vary the spin of the secondary. We provide a summary of these results in Table~\ref{table:summary_stats}.

\begin{table}[ht]
		\footnotesize
		\begin{center}
                        \setlength{\tabcolsep}{7pt} 
			\begin{tabular}{lc c c c c c|}
				\hline 
				    & Minimum & $Q1$ & Median & $Q3$ & Maximum \\ 
				\hline
				$\Delta{q}$ & -0.074 & -0.010 & -0.004 & 0.002 & 0.188  \\
				 \hline
				$\Delta{s_1}$  & -0.687 & -0.006 & 0.002 & 0.009 & 0.441 \\
				\hline
				$\Delta{s_2}$ & -0.466 & -0.010 & -0.000 & 0.009 & 0.767 \\
				\hline
				$\Delta{\sigma_{\textrm{eff}}}$ & -0.074 & -0.004 & 0.002 & 0.008 & 0.098 \\
				\hline
				$\Delta{S_{\textrm{eff}}}$ & -0.516 & -0.037 & -0.004 & 0.032 & 0.493 \\
				\hline
				$\cal{O}$ & 0.278 & 0.986 & 0.995 & 0.999 & 1.000 \\
			\end{tabular}
		\end{center}
	\caption{Summary of the distribution of absolute errors in the estimation of mass-ratio and individual spins over the entire parameter space under consideration: \(q\in[1,8],\, s_i\in[-0.8,\,0.8]\). We present minimum, median, maximum, first quartile, (\(Q1\)), and third quartile, (\(Q3\)), of the error distributions. Note that in addition to \((q,\,s^z_i)\), we also provide error summaries for the effective spin, \(\sigma_{\textrm{eff}}\), Hamiltonian effective-spin, \(S_{\textrm{eff}}\), and overlap between ground-truth signals and signals whose parameters are predicted by our network, \(\cal{O}\).}
	\label{table:summary_stats}
	\end{table}
	\normalsize

\noindent The last metric included in Table~\ref{table:summary_stats} corresponds to the overlap, ${\cal{O}}\left(h,\,s\right)$, between every waveform in the testing dataset, \(h(\theta^i)\) with ground-truth parameters \(\theta^i\rightarrow(q, s^z_i)\), and the signal, \(s\), that best describes \(h\) according to our neural network model, i.e., \(s(\hat{\theta}^i)\) with \(\hat{\theta}^i \rightarrow (q^\textrm{predicted}, s^{z,\,\textrm{predicted}}_i)\). It is necessary to consider this additional metric because absolute errors provide partial information about the accuracy of our neural network to infer the individual spins and mass-ratio of BBH mergers. One can only ascertain whether these predictions are reliable once we directly compare the signals described by the predicted parameters of our neural network model, and those that it is aiming to characterize. This is the theme of the following section.

\subsection{Quantitative Analysis}

\begin{figure}[h!]
\centerline{
\includegraphics[width=1.0\linewidth]{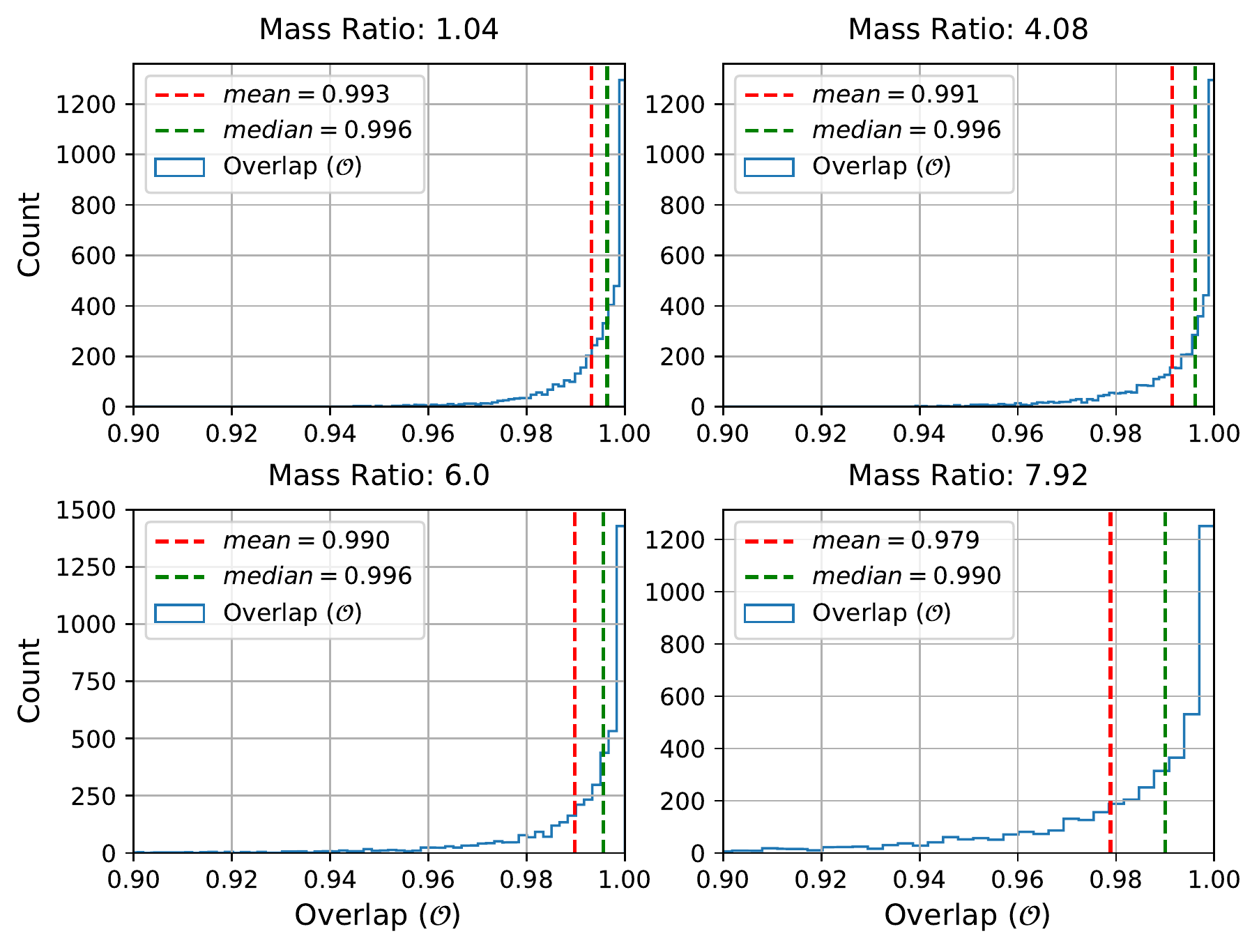}
} 
\caption{Histograms: Waveform Matches for a sample of test mass ratios}
\label{fig:overlap_hist}
\end{figure}

\noindent While the preceding qualitative analysis suggests that our neural network may be correctly inferring parameters for the signal manifold under consideration, we have also conducted a quantitative approach, i.e., we collected the predictions of the neural network for the mass-ratio and individual spins for each waveform in the testing data set. Thereafter, we generated waveforms with these \textit{predicted} parameters using \texttt{NRHybSur3dq8}. Finally, we computed the overlap between the ground-truth signals in the testing data set, and the waveforms whose parameters are predicted by our neural network. We summarize the results of this analysis for a sample of mass-ratio slices in Figure~\ref{fig:overlap_hist}. These results indicate that we are able to accurately infer the mass-ratio and individual spins over a broad range of the parameter space under consideration. Indeed, both the median and mean overlap results are above \({\cal{O}}\geq0.99\). They only drop below 0.98 for BBH mergers at the edge of the signal manifold with \(q\sim8\).

\begin{figure}[h!]
\centerline{
\includegraphics[width=1.0\linewidth]{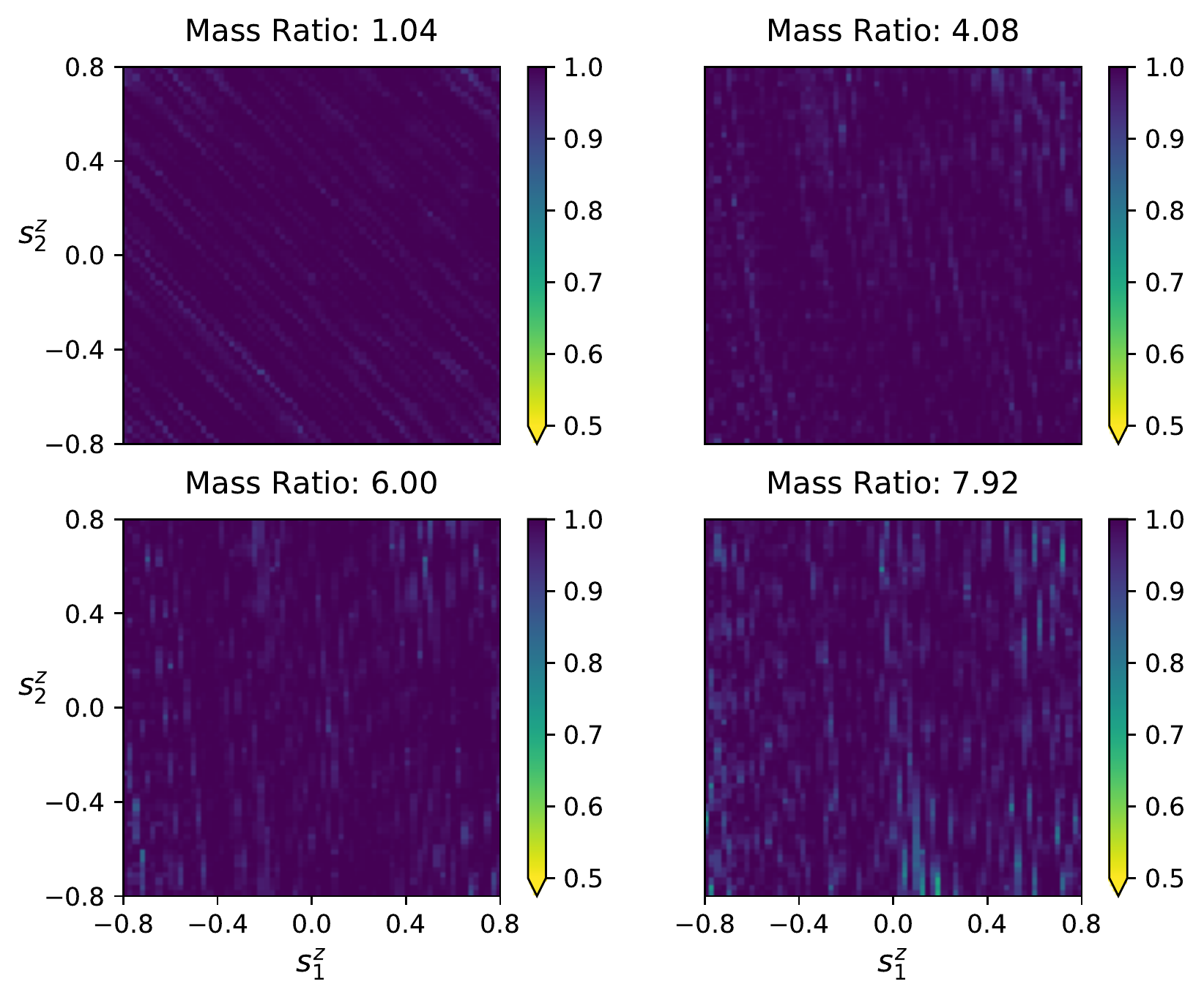}
} 
\centerline{
\includegraphics[width=0.55\linewidth]{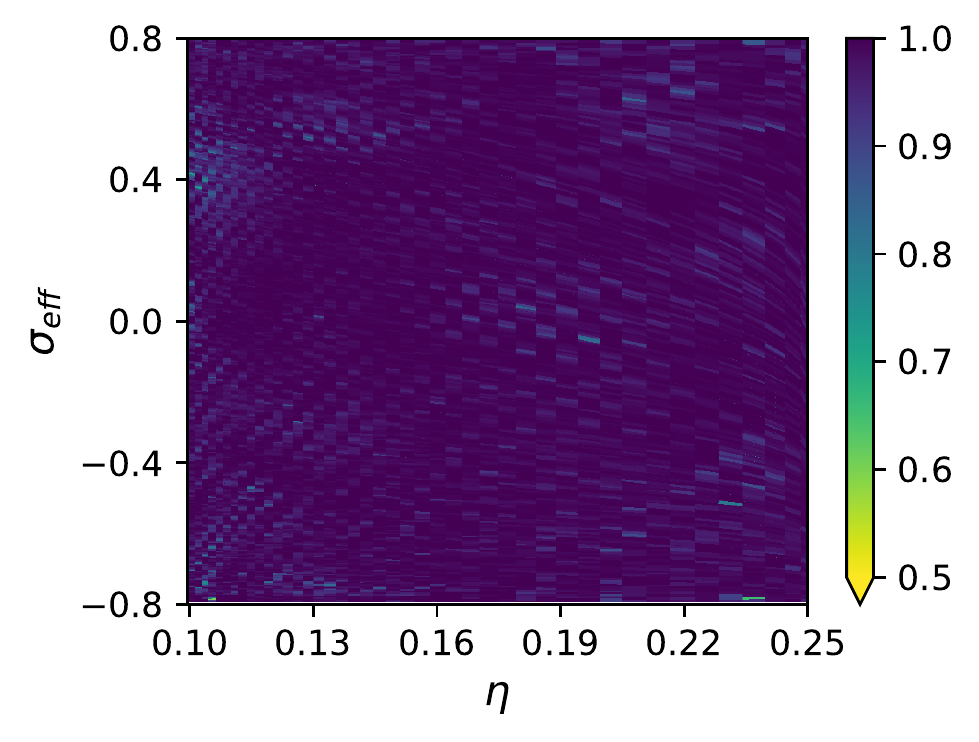}
} 
\caption{Each point in the top and middle panels represents the overlap between a signal in the testing data set and its counterpart whose individual spins and mass-ratio are predicted by our neural network model. The mass-ratio slices presented in this figure were randomly selected from the testing data set. The bottom panel summarizes the accuracy of our neural network model to infer the effective spin and symmetric mass-ratio,  \((\sigma_{\textrm{eff}},\, \eta)\), across the parameter space under consideration}.
\label{fig:wf_match_imshow}
\end{figure}

\noindent We provide a more detailed analysis of the overlap results in Figure~\ref{fig:wf_match_imshow}, where we show overlap results at every single point of the testing data set for a number of mass-ratio slices. These results show that our neural network model can predict the mass-ratio and individual spins of BBH mergers with excellent accuracy. Close to the edge of the parameter space under consideration, i.e., \(q\rightarrow8\), the neural network has a gradual decrease in accuracy. The bottom panel in Figure~\ref{fig:wf_match_imshow} also presents results for the accuracy with which our neural network model is able to reconstruct the effective spin, \(\sigma_{\textrm{eff}}\), and symmetric mass-ratio, \(\eta\), of BBH mergers across the parameter space under consideration.

\begin{figure}[h!]
\centerline{
\includegraphics[width=0.33\linewidth]{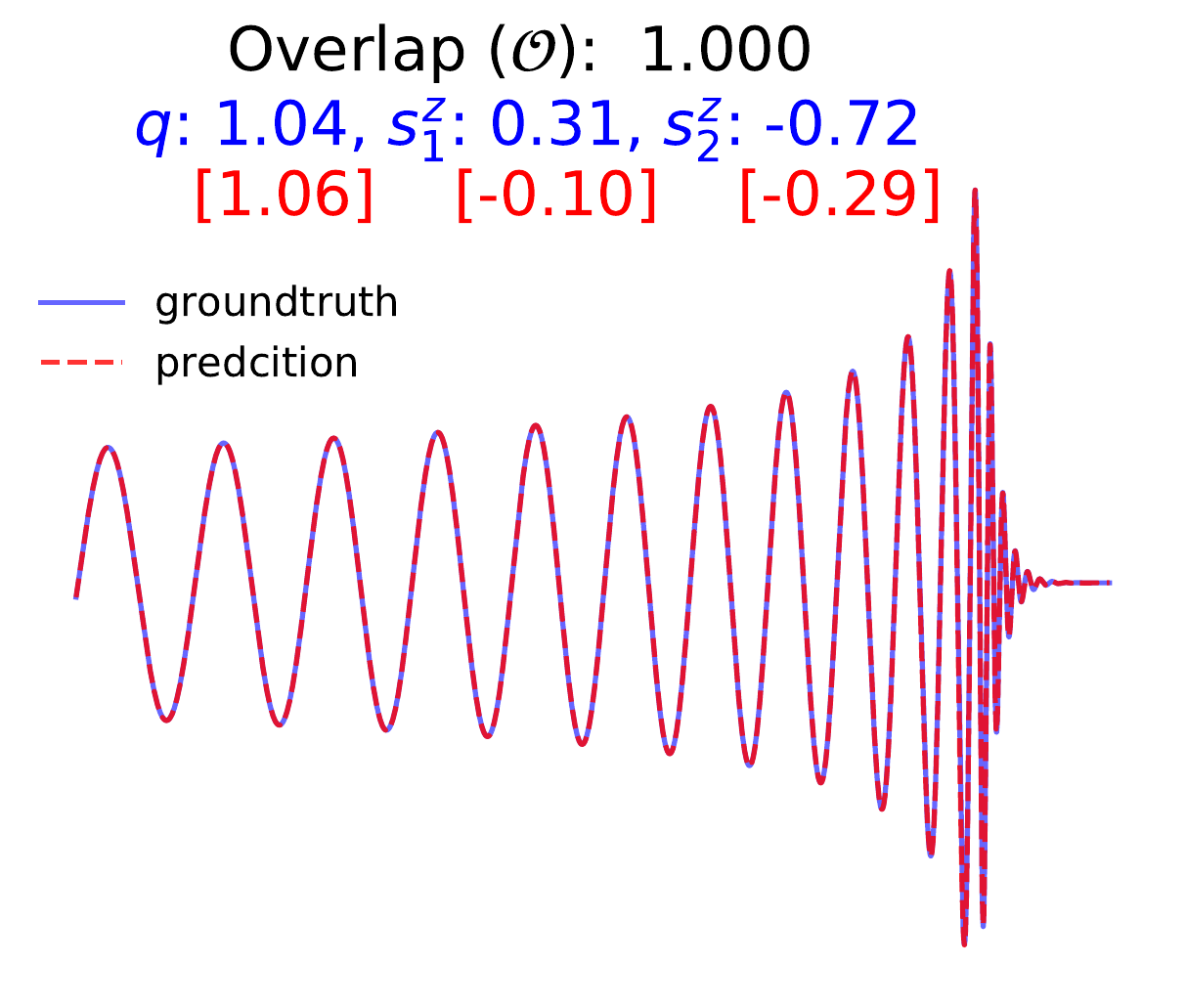}
\includegraphics[width=0.33\linewidth]{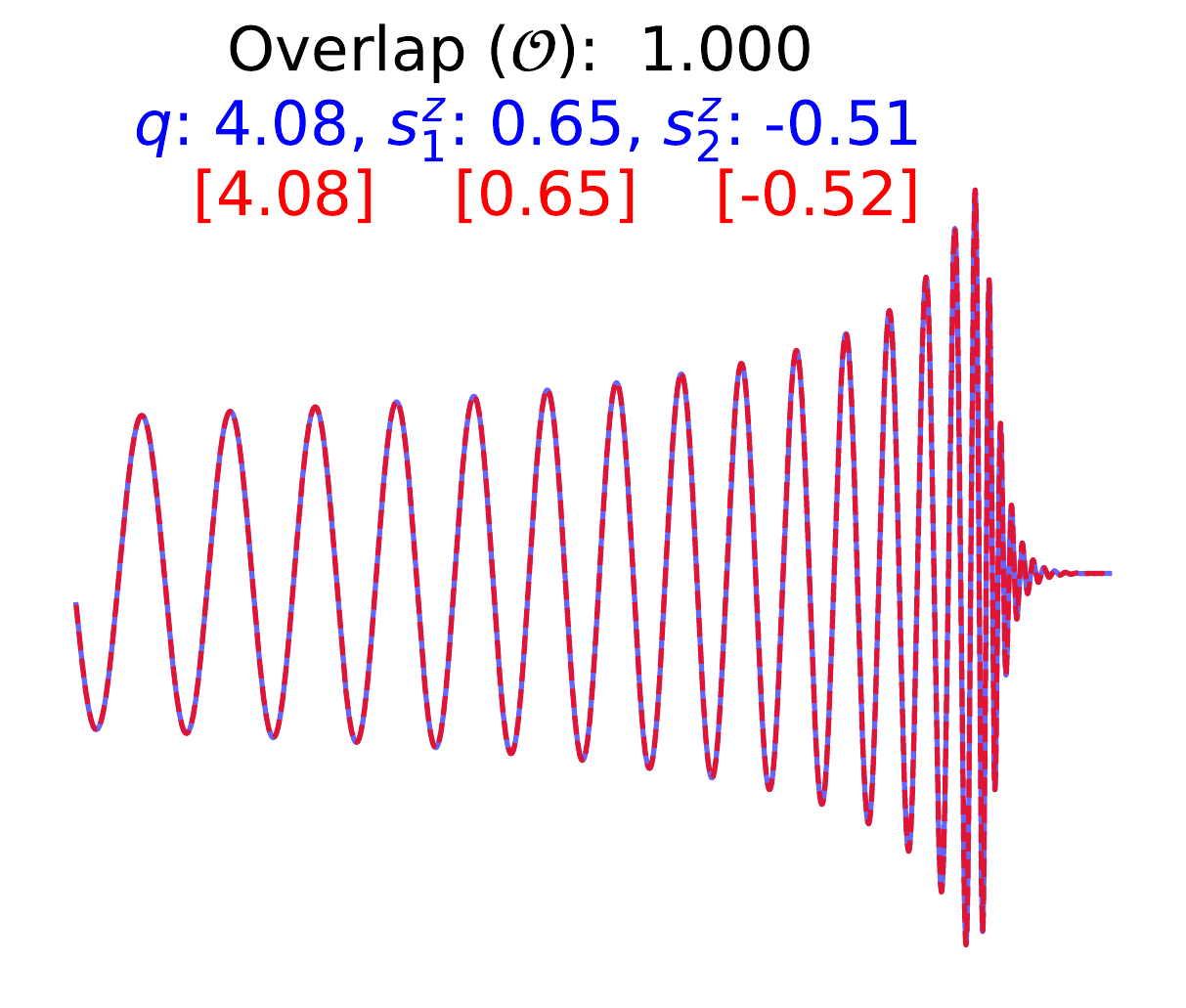}
\includegraphics[width=0.33\linewidth]{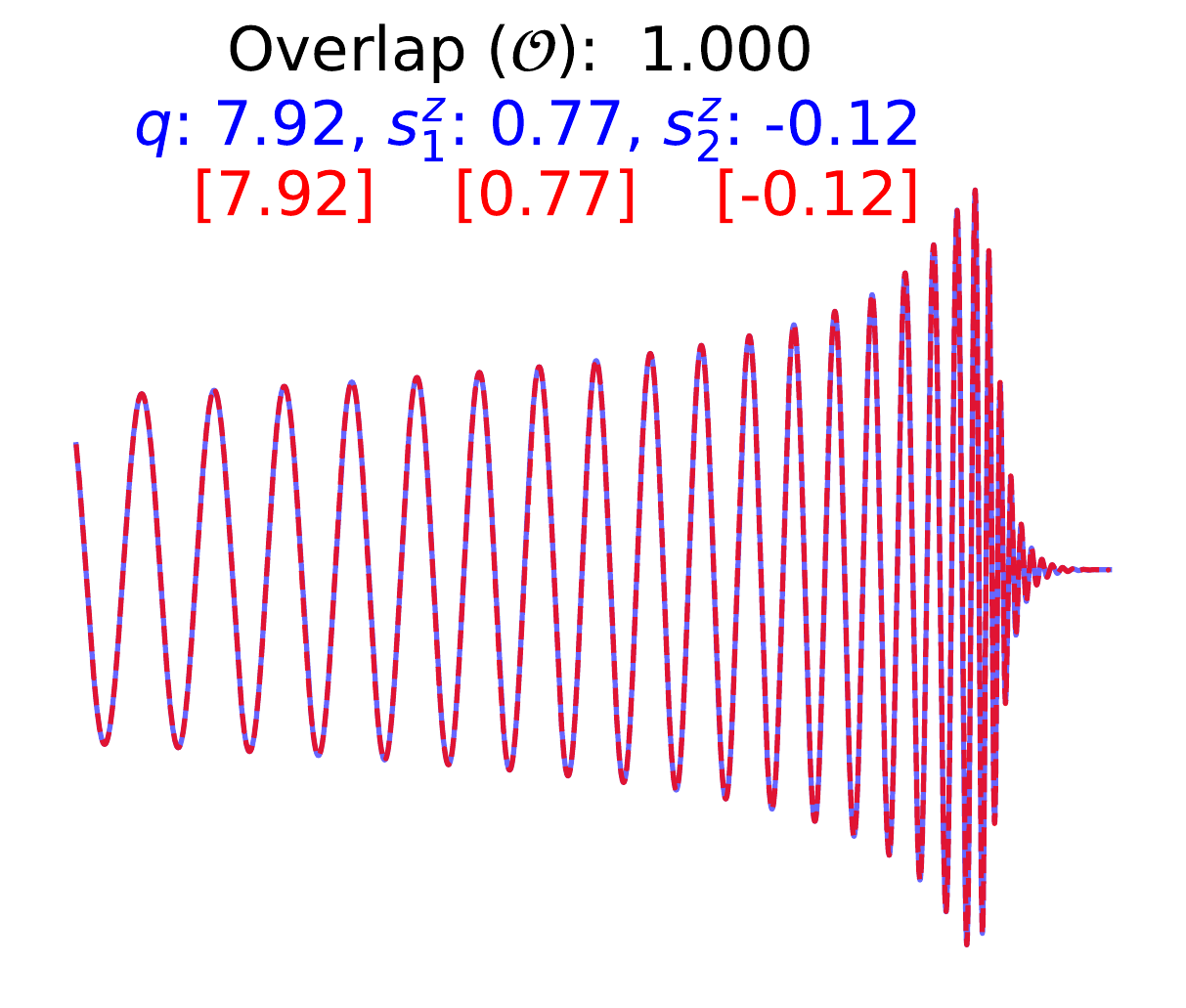} }
\centerline{
\includegraphics[width=0.33\linewidth]{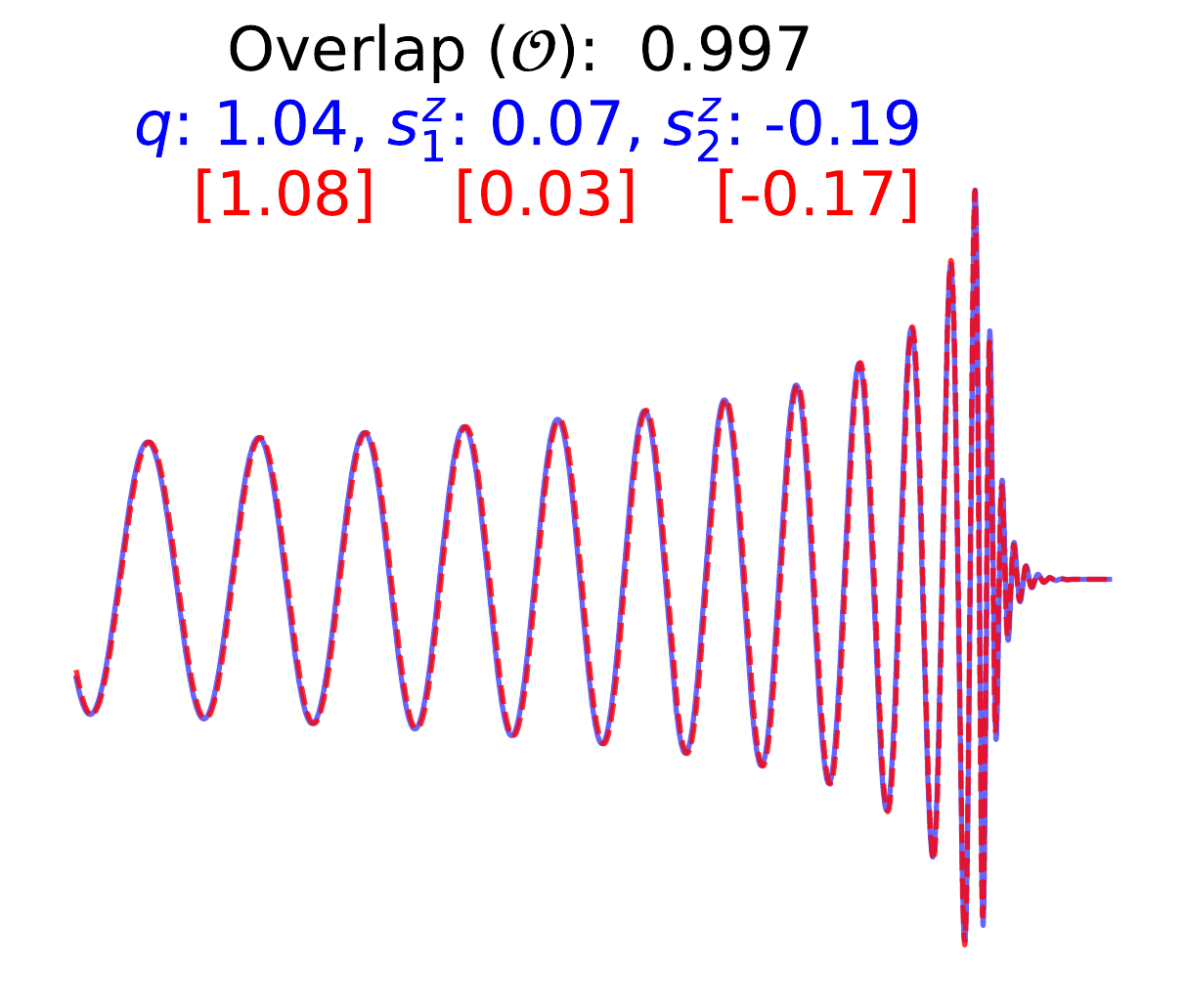}
\includegraphics[width=0.33\linewidth]{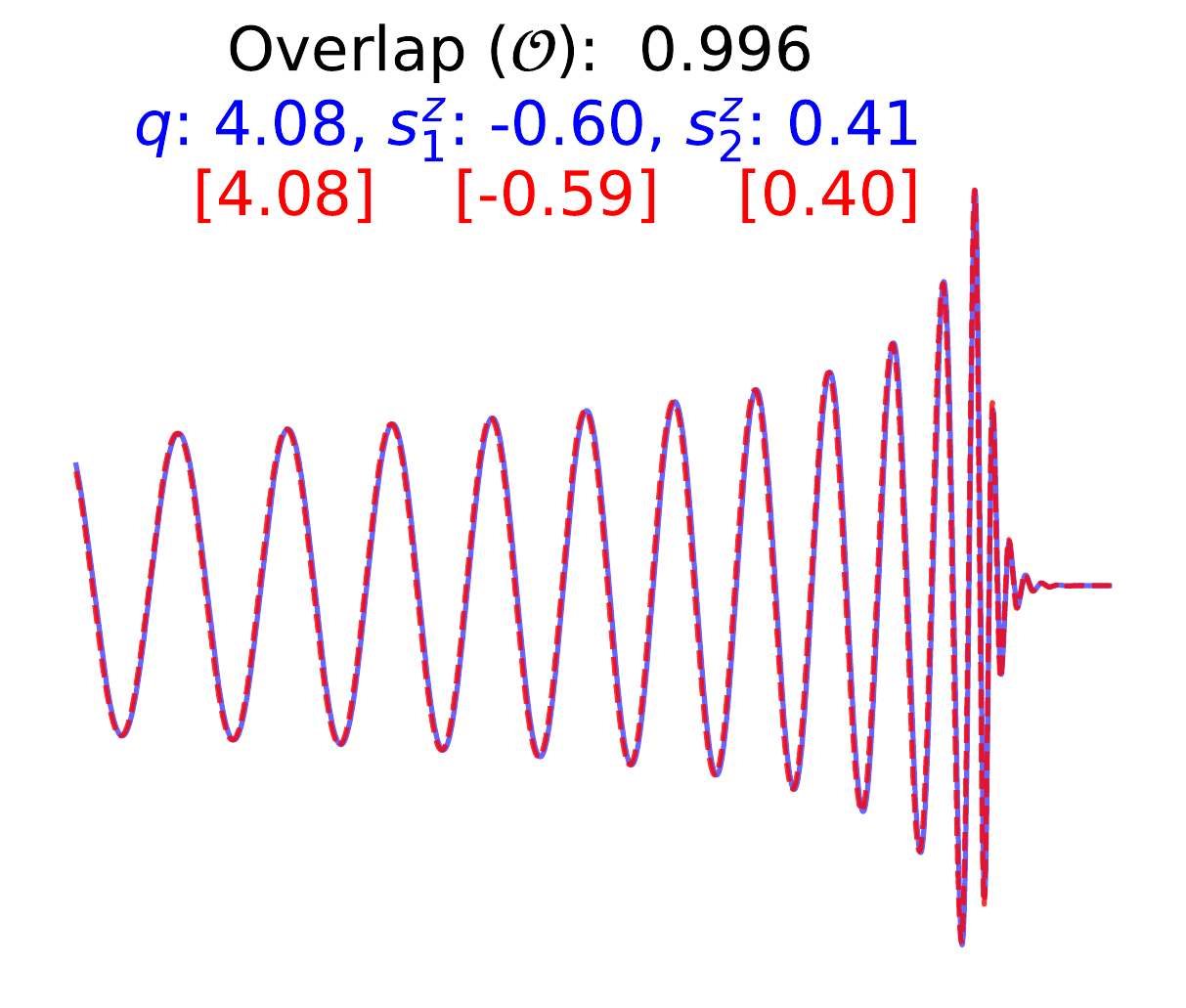} 
\includegraphics[width=0.33\linewidth]{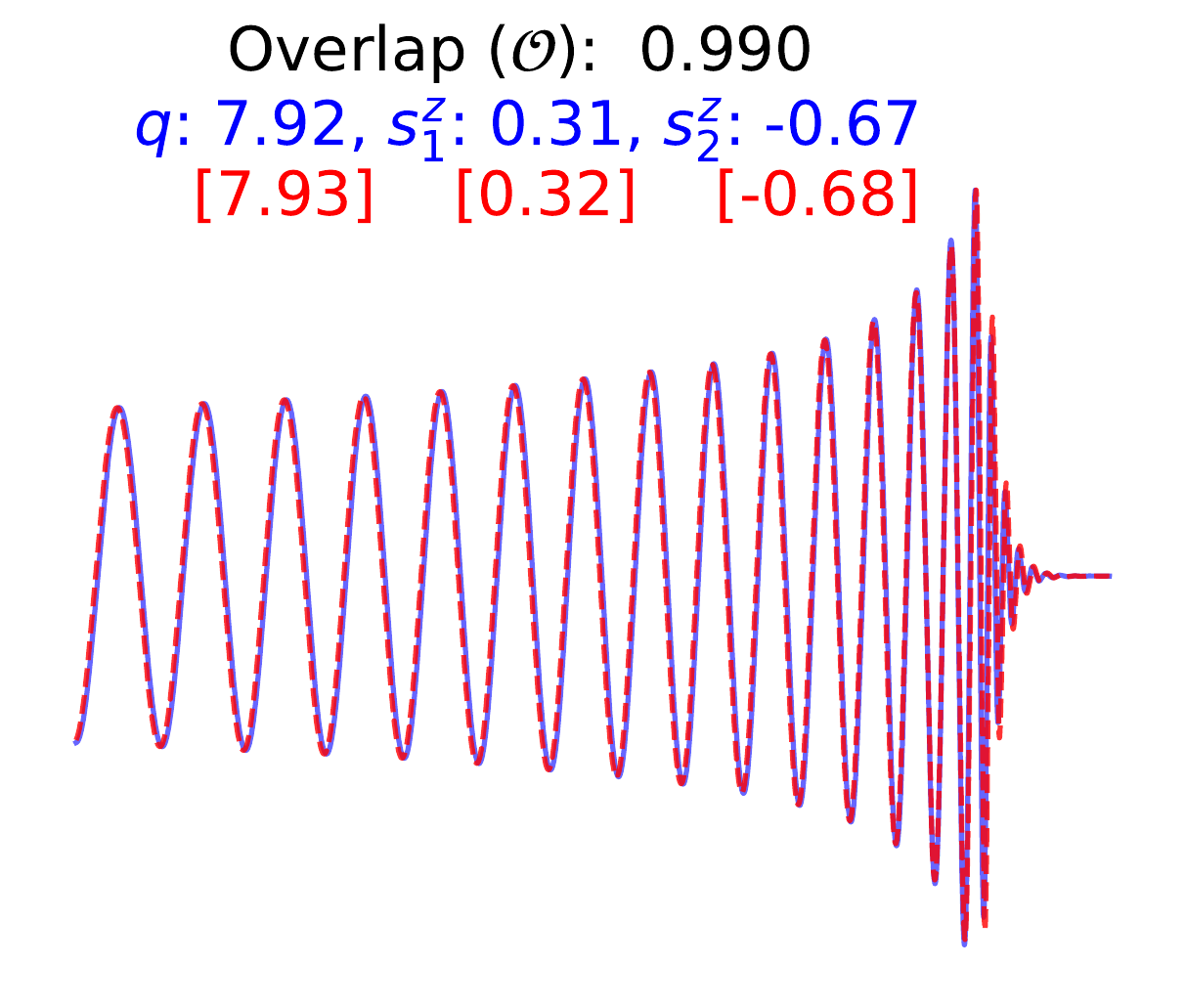}
} 
\centerline{
\includegraphics[width=0.33\linewidth]{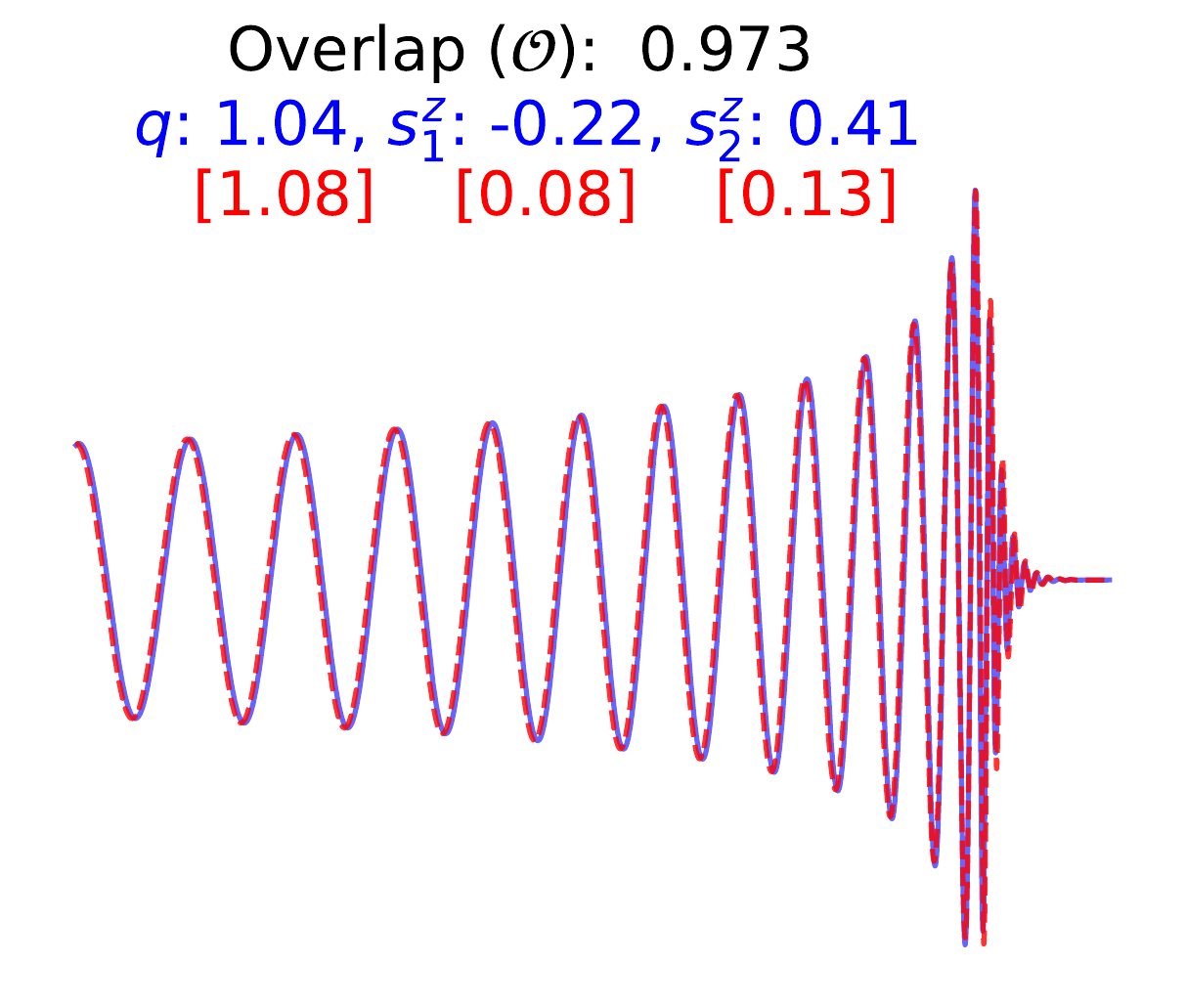}
\includegraphics[width=0.33\linewidth]{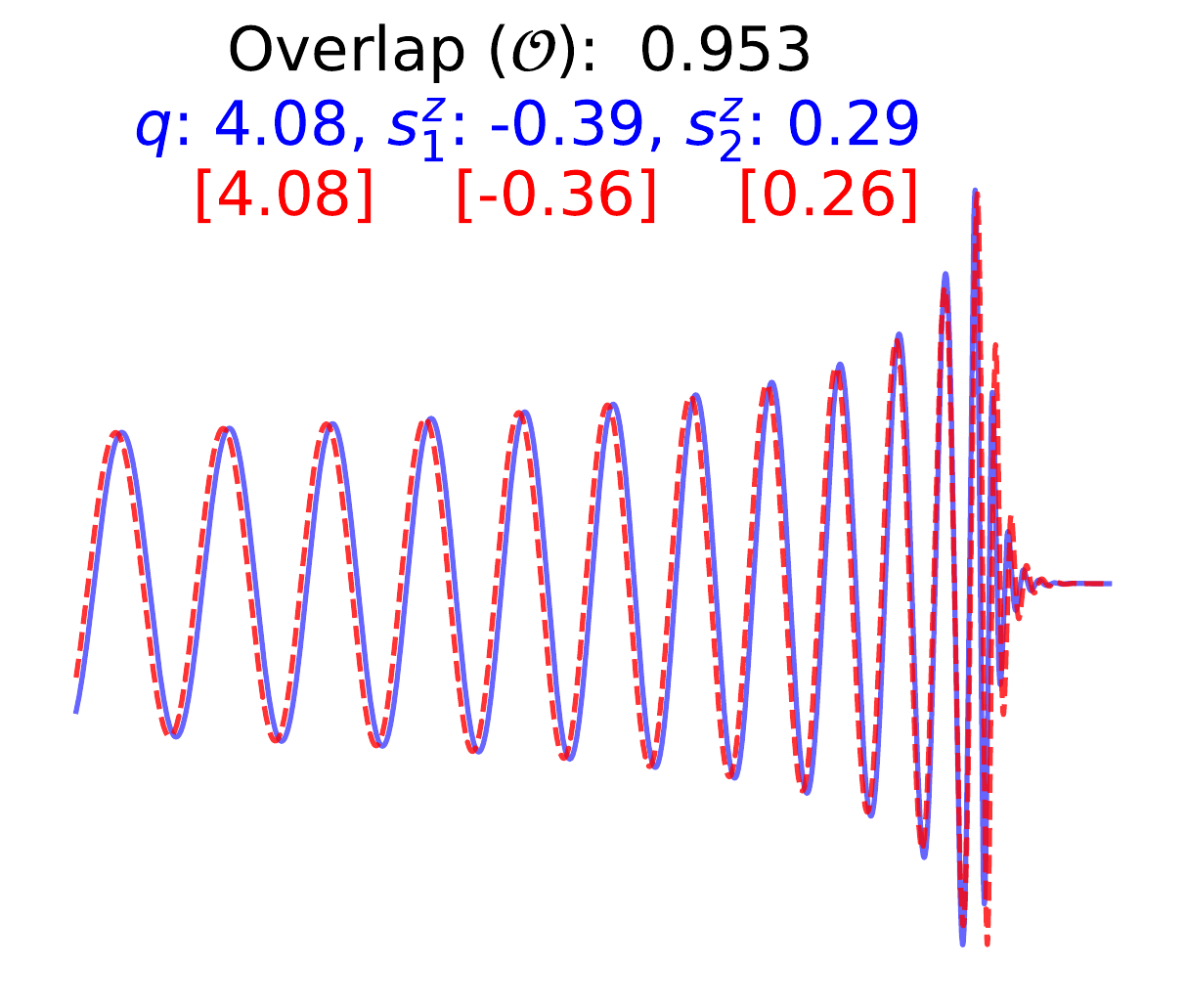} 
\includegraphics[width=0.33\linewidth]{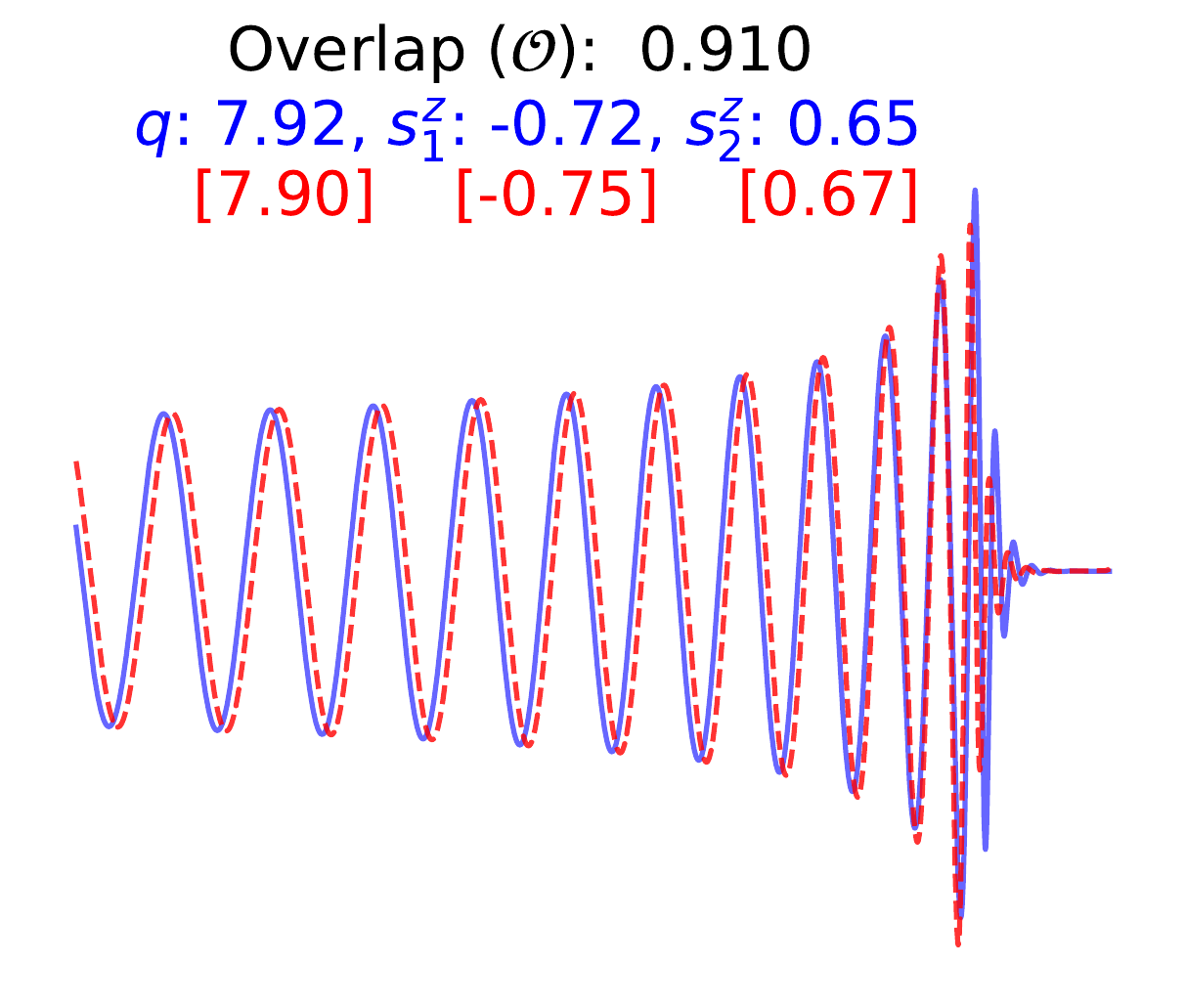}
} 
\caption{Random sample of high (top panel), median (middle panel) and low (bottom panel) overlap $\cal{O}$ predictions for $q=$ 1.04 (left column), 4.08 (middle column), 7.92 (right column). Low overlaps are two sigma below mean overlap for each $q$.}
\label{fig:wf_match_samples}
\end{figure}

\noindent Figure~\ref{fig:wf_match_samples} presents a visual representation of high, median and low overlaps samples, i.e., 2 sigma below the median overlap. These random samples from the testing data set show that: (i) our neural network can identify signals that reproduce with excellent accuracy the dynamics of near-equal BBH mergers. However, given the parameter space degeneracy of the signal manifold for \(q\sim1\) systems, it is difficult to accurately recover the individual spins of these systems---see left column in Figure~\ref{fig:wf_match_samples}; (ii) for BBH systems with \(q\neq 1\)---middle column in Figure~\ref{fig:wf_match_samples}---we notice that our network can recover with excellent accuracy both the mass-ratio and individual spins; and (iii) systems with asymmetric mass-ratios---right column in Figure~\ref{fig:wf_match_samples}---can be characterized with different levels of accuracy. We notice that even for systems whose overlap is two sigma below the median overlap, the neural network is able to tightly constrain the estimate of the individual spins of the systems, as already summarized in Table~\ref{table:summary_stats}.

\noindent The quantitative analysis we have conducted indicates that deep learning is adequate to reconstruct the mass-ratio and individual spins of NR-type time-series signals that span the 3-D signal manifold \(q\in[1,\,8]\) and \(s_{\{1,\,2\}}^z\in[-0.8,\,0.8]\). Our findings are in good accord with the expectation that it is hard to estimate the individual spins of BBH mergers whose mass-ratios are near the edges of the signal manifold, i.e., \(q\sim1\) and \(q\sim 8\). This is expected given the symmetry of the problem at hand for \(q\sim1\), and the fact that the spin of the secondary is difficult to reconstruct for asymmetric mass-ratio BBH mergers. All these results may be interactively perused at~\cite
{interactive}.

We provide a careful follow up of BBH systems that are not accurately reconstructed by our neural network in~\ref{sec:ap1}, which tend to be concentrated at the edges of the signal manifold.

While we have found that our deep learning model performs very well to characterize \(q\sim1\) BBH mergers, as shown in the top left panel of Figure~\ref{fig:overlap_hist}, we leave to future work the construction of deep learning models that include higher-order modes to explore whether spin reconstruction may be improved for the most asymmetric mass-ratio BBH mergers considered in this study.

\section{Conclusion}
\label{sec:end}

Inferring the spin distribution of BBH mergers will unveil new and detailed information about the formation channels of these objects. It is then timely and relevant to design signal-processing algorithms that are scalable, and which may readily handle a dense sampling of this higher-dimensional signal manifold.

To contribute to this effort, in this paper we introduced a deep learning model to estimate the individual spins, effective spin and mass-ratio of quasi-circular, spinning, non-precessing, BBH mergers. To do this, we densely sampled the \((q,\, s_1^z,\, s_2^z)\) parameter space with over 1.5M waveforms produced with the \texttt{NRHybSur3dq8} model. We then designed and deployed a distributed training algorithm to reduce the training stage from one month, using a single V100 GPU, to just 12.4 hrs using 64 V100 GPUs in the HAL cluster at the University of Illinois. The convergence of deep learning and high performance computing proved critical to reduce time-to-insight, and enabled us to test a variety of optimization algorithms that incorporate general relativistic constraints. This approach enabled us to demonstrate that, in the absence of noise, physics-inspired deep learning models can effectively reconstruct the mass-ratio and individual spins of the binary components of BBH systems. 

Our study sheds light on the ability of deep learning to characterize waveform signals that span a degenerate signal manifold. For instance, in the case of equal or comparable mass-ratio systems we showed that our neural network can reconstruct the mass-ratio of these systems with excellent accuracy. However, the intrinsic symmetry of these systems, i.e., the binary components are indistinguishable \(m_1\leftrightarrow m_2\), reduces the ability of the neural network to provide good estimates of the individual spins of the system. Despite these difficulties, we have found that the signals predicted by our network reproduce the dynamics of the ground-truth signals with excellent accuracy, i.e. we observe a median overlap of $99.6\%$ between the predicted and the ground-truth signals for near equal mass ratios. On the other hand, the neural network can tightly constrain both the mass-ratio and individual spins for asymmetric mass-ratio systems. Near the edge of the signal manifold, i.e.,  \(q\sim8\) BBH mergers, the neural network has a drop in accuracy recovering the spin of the secondary. This is expected because the spin of the secondary has a less dominant effect in the overall dynamics of the waveform signal. We leave to future work a systematic assessment of the importance of including higher-order modes to better constrain the spin of the secondary. 

In summary, these results represent a significant step towards the design of scalable, physics-inspired neural network models that may be used to infer the spin distribution of actual GW observations. This analysis will enable a firm understanding on the impact of real GW noise when deep learning is used to infer the spin distribution of BBH mergers. An analysis of this nature will be presented in a forthcoming publication.

\section{Acknowledgements}
\label{ack}
We thank Arjun Shankar, Tom Gibbs, Junqi Yin, and Jeff Larking for their support and guidance using the Summit supercomputer. We also thank Ben Blaiszik, Ryan Chard and Logan Ward for their support deploying our neural network model and testing dataset at the Deep Learning Hub hosted by Argonne National Lab.

We thank Elise Jennings for her guidance and support using the Cooley and Theta supercomputers at Argonne National Lab. 

AK and EAH gratefully acknowledge National Science Foundation (NSF) awards OAC-1931561 and OAC-1934757. We are grateful to NVIDIA for donating several Tesla P100 and V100 GPUs that we used for our analysis. AD acknowledges support from the National Center for Supercomputing Applications (NCSA) Students Pushing INnovation (SPIN) undergraduate internship program. The authors thank Dawei Mu and Chit Khin for their superb support using the HAL cluster and the ICCP, respectively.

This work utilized resources supported by the NSF's Major Research Instrumentation program, the Hardware-Learning Accelerated (HAL) cluster, grant OAC-1725729, as well as the University of Illinois at Urbana-Champaign. This research used resources of the Oak Ridge Leadership Computing Facility, which is a DOE Office of Science User Facility supported under Contract DE-AC05-00OR22725.

This work made use of the Illinois Campus Cluster, a computing resource that is operated by the Illinois Campus Cluster Program (ICCP) in conjunction with the NCSA and which is supported by funds from the University of Illinois at Urbana-Champaign. This research used resources of the Argonne Leadership Computing Facility, which is a DOE Office of Science User Facility supported under Contract DE-AC02-06CH11357. We also acknowledge NSF grant TG-PHY160053, which provided us access to the Extreme Science and Engineering Discovery Environment (XSEDE) Bridges-AI system. We thank the \href{http://gravity.ncsa.illinois.edu}{NCSA Gravity Group} for useful feedback. 
\bibliography{ecc_references}
\bibliographystyle{apsrev4-1}

\appendix

\clearpage
\section{Overlap distribution}
\label{sec:ap1}

We have conducted a detailed analysis of the regions of parameter space where our neural network model does not perform optimally. By the numbers, the test data set has 197,516 waveforms. Our neural network predicts signals whose overlap with the corresponding ground-truth signals in the test set has the following distribution:

\begin{cititemize2}
\item Test waveforms with \({\cal{O}}\leq0.97\): 18,315
\item Test waveforms with \({\cal{O}}\leq0.90\): 817
\item Test waveforms with \({\cal{O}}\leq0.85\): 181
\end{cititemize2}

\begin{figure}[h!]
\centerline{
\includegraphics[width=0.5\linewidth]{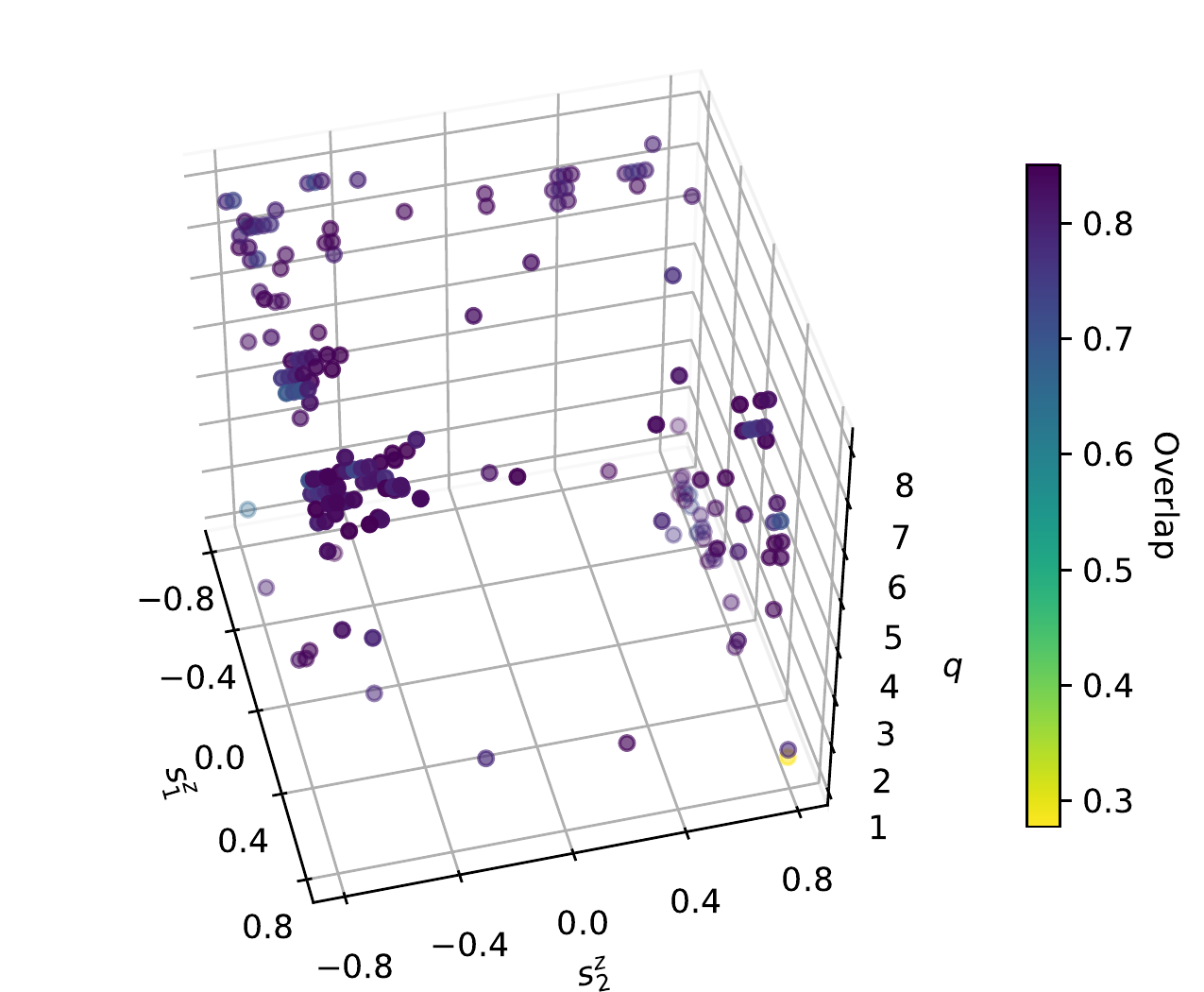}
\includegraphics[width=0.5\linewidth]{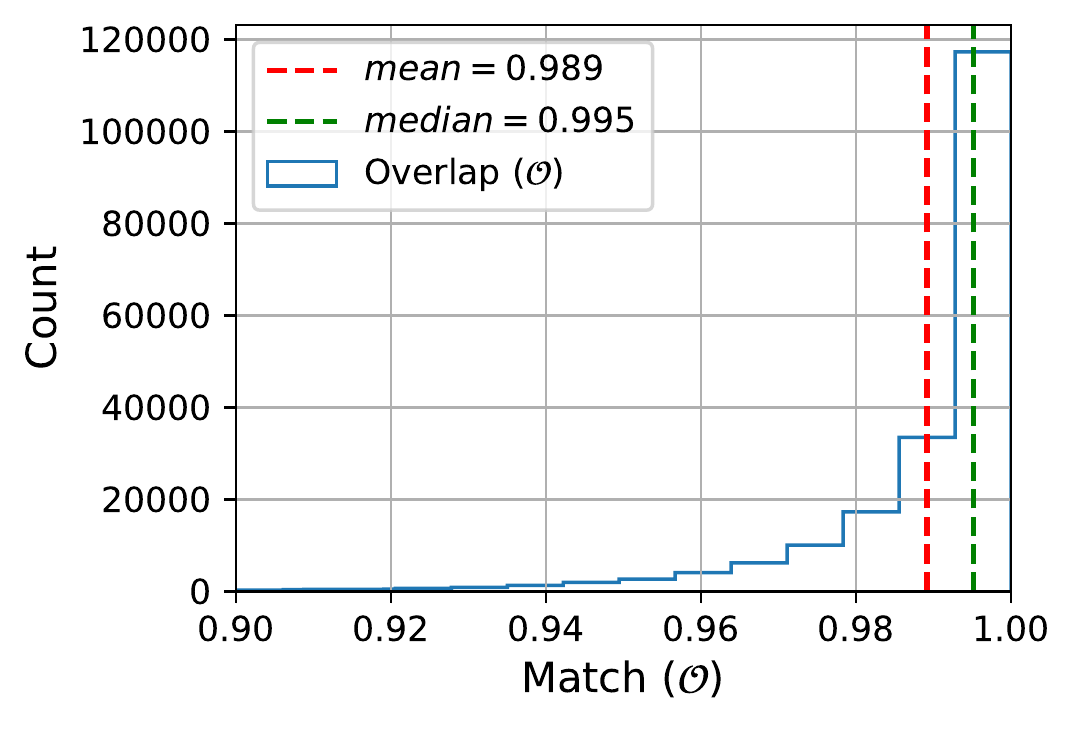}
} 
\caption{Left: All test set points with Overlap $< 0.85$. Right: overlap distribution for the entire test set.}
\label{fig:Low_Overlaps}
\end{figure}

These results are presented at a glance in Figure~\ref{fig:Low_Overlaps}. We have also found that if we excise the edges of the signal manifold, i.e., we consider the ranges  \(q\in[1,\,8]\), \(s_{\{1,\,2\}}\in(-0.8,\,0.8)\), then most of the low overlaps are removed. 

\noindent The reader may conduct a more detailed inspection of these results at~\cite
{interactive}. Furthermore, as mentioned in the main body of this article, it is worth exploring whether the inclusion of higher-order modes may  improve the performance of this model for asymmetric mass-ratio BBH mergers. Such a study will be pursued in the near future.

\end{document}